\documentclass[reprint,amsmath,amssymb,aps]{revtex4-2}

\usepackage{hyperref}
\hypersetup{
    colorlinks=true,
    linkcolor=blue,
    citecolor=blue,
    urlcolor=blue
}

\usepackage{graphicx}

\usepackage{mathdots}
\usepackage{xcolor}

\newcommand{\dbar}[1]{\overline{\overline{#1}}}

\usepackage[toc,page]{appendix}

\begin{document}

\title{Generalized Huygens' condition as the fulcrum of planar nonlocal\\omnidirectional transparency: from meta-atoms to metasurfaces}

\author{Amit Shaham}
\email{samitsh@campus.technion.ac.il}

\author{Ariel Epstein}
\email{epsteina@ee.technion.ac.il}

\affiliation{Andrew and Erna Viterbi Faculty of Electrical and Computer Engineering, Technion -- Israel Institute of Technology, Haifa 3200003, Israel}

\begin{abstract}
{Fresnel reflection has been known for centuries to fundamentally impede efficient transmittance across planar interfaces, especially at grazing incidence. Herein, we present the generalized Huygens' condition (GHC) to resolve such intricacies in metasurface (MS) designs and allow omnidirectional transparency. Compared to common numerical antireflective-coating approaches, our analytical framework yields surprisingly simple closed form conditions, carefully leveraging the natural nonlocal mechanisms endowed in planar electromagnetic structures. At the meta-atom level, we meet the GHC by balancing traditional tangential susceptibilities of Huygens' MSs with their unconventional normal counterparts; the latter facilitates the key requisite of vanishing backscattering at the challenging grazing incidence scenario. At the MS level, we sheerly utilize this central insight to engineer realistic all-angle transparent printed-circuit-board (PCB) cascaded admittance sheets. Thoroughly validated in simulation and experiment, this universal GHC demonstrates a resourceful venue for practical implementation of advanced nonlocal devices, e.g., flat optical components, optical analog computers, and spaceplates.
}
\end{abstract}

{\hypersetup{hidelinks}
\maketitle
}

\section{Introduction}
\label{Sec:Intro}
Fresnel reflection \cite{Fresnel1823} is a fundamental phenomenon of waves, such as light, electromagnetic radiation, and sound \cite{Jenkins1950,Jackson1999,Balanis2012,Pozar2012,Lekner2016}. Although vital for many everyday and scientific utilities, e.g., mirrors, interferometers, and waveguides, it often poses an undesired nuisance, as it impairs the performance of other transmissive and absorptive applications, e.g., lenses, protective antenna radomes, absorbers, energy harvesters, and photovoltaic cells.

For planar interfaces, classical antireflective coatings of various refractive-index profiles \cite{Rayleigh1879,Chattopadhyay2010,Raut2011,Pozar2012,Balanis2012,Lekner2016} allow one to shape the spectral and angular range of low reflectance. Modern designs, e.g., \cite{Zhang2009,Chen2010,Chattopadhyay2010,Raut2011,Chu2021}, utilize artificial metamaterials to relax thickness constraints or to improve other figures of merit. Such evolution has invoked the ultimate goal of \emph{all-angle} antireflection \cite{Chattopadhyay2010,Raut2011}. To this end, antireflective-coating and metamaterial approaches, e.g., \cite{Dobrowolski2002,Poitras2004,Im2018}, have numerically optimized profiles of refractive-index values; other schemes have avoided the intricacies involved in planar Fresnel reflection, for instance, by meshing engineered invisible inclusions of spherical or cylindrical geometries (whose all-angle transparency is naturally guaranteed by their symmetry) into structured bulks \cite{Alu2007,Alu2009,Padooru2012,Ye2016}.

One major challenge in such reflectionless planar designs is the inherent angular dependence of Fresnel reflection. In view of backscattering cancellation as a problem of impedance matching \cite{Pozar2012,Balanis2012}, the wave impedance in each medium strongly depends on the angle of incidence, and, particularly, approaches extreme values at grazing incidence (infinite or zero, depending on the polarization) \cite{Lekner2016}; hence, it typically leads to considerable reflectance in the overall angular behavior, even at more acute angles.

In fact, almost all the aforementioned remedies rely on exhaustive numerical or full-wave optimizations, which many times hinder intuitive physical interpretations, qualitative trends, and knowledge regarding the very existence of a feasible solution, let alone its optimality. For example, the inevitable singular impedance values (unity reflectance) at grazing incidence have been previously noticed \cite{Dobrowolski2002,Poitras2004}, yet circumvented via numerical optimizations at other angles; seemingly, no fundamental investigation as to their origin or solutions to directly counter this long standing issue have been reported so far.

Interestingly, a hint for a simple viable solution turns out to reside in a different engineering paradigm that utilizes another elementary property of waves: Huygens' principle \cite{Huygens1690,Love1901}, or, in its modern form, Schelkunoff's surface equivalence theorem  \cite{Schelkunoff1936,Balanis2012,Pozar2012,Balanis2016}. Recently, such equivalence found extensive use in modern wave-manipulating devices called metasurfaces (MS) \cite{Glybovski2016,Chen2016}.

In general, MSs are sheets of subwavelength thickness accommodating dense arrays of artificial scatterers. They have demonstrated extraordinary capability of wavefront, scattering, and polarization shaping, both in microwave and optical frequencies \cite{Glybovski2016,Chen2016}. As a fundamental example, collocated tangential electric and magnetic responses have been shown to deflect or focus an incident beam with low reflectance \cite{Pfeiffer2013Huygens,Monticone2013,Selvanayagam2013}; such devices have been termed ``Huygens' metasurfaces'' (HMS), after Huygens' principle. Later on, low-profile printed-circuit-board (PCB) versions have emerged \cite{Pfeiffer2013Cascaded,Pfeiffer2013Millimeter,Wong2014}, facilitating low-cost fabrication via standard and widely available techniques.

At the conceptual \emph{meta-atom level} of the design \cite{Pfeiffer2013Huygens,Selvanayagam2013,Pfeiffer2013Millimeter,Wong2014}, each scatterer placed on a HMS is engineered to satisfy Huygens' condition, namely, to 
locally introduce balanced electric and magnetic dipole moments, typically at normal incidence. Subject to this condition, their back-scattered waves cancel each other \cite{Love1976,Ziolkowski2010,Decker2015,Ziolkowski2022}, while their forward-scattered waves interfere to yield desirable local phase shift. This, in fact, serves as another footing to the impedance matching approach, providing intuitive physical interpretation and a closed-form framework. At the \emph{MS level} of design, a more practical structure, such as a PCB implementation of admittance-sheet cascades \cite{Pfeiffer2013Cascaded,Pfeiffer2013Millimeter,Monticone2013}, is usually considered to effectively realize such dipole responses. Huygens' condition is then effectively emulated by tuning the constitutive parameters of the composite (e.g., the surface admittance of each layer on the PCB), typically with the aim to equalize excitation of the fundamental symmetric (electric) and antisymmetric (magnetic) modes supported by the apparatus \cite{Pfeiffer2013Millimeter,Decker2015}.

While HMSs have been successfully integrated in many applications \cite{Pfeiffer2013Millimeter,Wong2014,Epstein2016,EpsteinNature2016,Chen2018}, they still suffer a fundamental limitation in the spirit of Fresnel reflection. Namely, for MSs of  \emph{tangentially polarizable} inclusions, Huygens' condition is inherently angularly dependent in the sense that it applies solely to a single predefined angle of incidence (typically normal), while its unity transmittance wanes as deviation from this angle is increased \cite{Holloway2005,Epstein2016}. This angular or wavevector sensitivity indicates a more subtle and universal mechanism at work: spatial dispersion or nonlocality \cite{Overvig2022,Shastri2023}, where the response of a system (e.g., a MS) at a certain location depends not only on the input fields at the observation point, but on the fields at different locations as well. Equally as well, the device response to an incoming wave at a given point depends not only on the amplitude of the field \emph{in situ}, but also on the angle of incidence, or, more generally, on the global phase profile or field distribution in space.

Recently, nonlocality has regained immense interest as a powerful tool to surpass local applications \cite{Overvig2022,Shastri2023}, facilitating, \emph{inter alia}, angular filters \cite{Mailloux1976,Franchi1983,Ortiz2013,Shaham2022}, wide-angle impedance matching layers \cite{Magill1966,Cameron2015}, HMSs with extended functionalities \cite{Epstein2016PRL,Pfeiffer2016Emulating}, flat nonlocal optical elements and lenses \cite{Genevet2017}, and other advanced operations, such as space-squeezing plates \cite{Guo2020,Reshef2021} and optical analog computing \cite{Silva2014,Kwon2018,Abdollahramezani2020,Momeni2021}. At the meta-atom level, local polarizable elements can emulate spatially dispersive response by channeling the intrinsic nonlocality imbued in Fresnel scattering or Maxwell's equations (due to the spatial derivatives relating between field components). One effective way to achieve this is by introducing \emph{normally polarizable} inclusions \cite{Ortiz2013,Achouri2015,Cameron2015,Zaluski2016,Pfeiffer2016Emulating,Albooyeh2017,Achouri2017,delRisco2021,Momeni2021,Shaham2021,Shaham2022} in addition to the standard tangentially polarizable ones and thus allow more degrees of freedom to course such natural nonlocality. Nevertheless, this useful technique has been widely overlooked since most MSs are utilized for single-angle functionalities, for which tangential elements typically suffice \cite{Albooyeh2017}.

In this paper, we offer a comprehensive solution to overcome these challenges by deriving the generalized Huygens' condition (GHC): extending the original single-angle concept of Huygens' condition \cite{Pfeiffer2013Huygens,Monticone2013,Selvanayagam2013,Pfeiffer2013Cascaded,Pfeiffer2013Millimeter,Wong2014,Epstein2016}, we allow all-angle transparency, both at the meta-atom and MS levels. Following the generalized sheet transition conditions (GSTCs) \cite{Idemen1990,Tretyakov2003,Kuester2003,Achouri2015}, we surprisingly find that such a functionality merely requires the satisfaction of two simple closed-form conditions: zero reflection at normal and grazing incidence scenarios (Sec.\ \ref{Sec:GHC}). At the meta-atom level, we utilize the unique nonlocal properties of \emph{normally polarizable} inclusions, alongside tangential responses, to realize the GHC. In particular, we reveal that such normal susceptibilities are the key to enforcing vanishing reflection at \emph{grazing} incidence; we thus, in fact, resolve the elusive issue of total reflection therein, which has hitherto been deemed immutable \cite{Dobrowolski2002,Poitras2004}, and stabilize the angular behavior of Huygens' condition. Subsequently, we validate the GHC by devising and tuning a rigorous physical set of compound meta-atoms via full-wave simulations (Sec.\ \ref{Sec:MetaAtom}).

Inspired by these GHC-related observations, we proceed to a practical MS-level design of thin PCB cascades of admittance sheets separated by a dielectric substrate (Sec.\ \ref{Sec:MSLevel}). Following standard transmission-line (TL) formalism, we analyze the scattering off such a configuration and stipulate zero reflection at grazing incidence, in contrast to the typical Huygens' condition at normal incidence \cite{Pfeiffer2013Huygens,Monticone2013,Selvanayagam2013,Pfeiffer2013Cascaded,Pfeiffer2013Millimeter,Wong2014,Epstein2016}. We arrive at a simple unique closed-form solution to satisfy this condition and show that the resultant composite is essentially transparent at all angles, both by transmitted magnitude and phase. Beyond observing the excellent performance exhibited by these designs, both in simulation and experiment, we demonstrate that the MS-level design is fundamentally related to the meta-atom level by satisfying effective susceptibility balance at grazing incidence. This covenant captures the intricate properties of the diverse nonlocal phenomena governing such configurations and indicates intriguing potential to emulate normal susceptibilities by stacking simple tangential components. Overall, our results verify that the GHC can indeed be harnessed as a fundamental principle to implement reflectionless nonlocal functionalities at the entire angular range. In the future, this evidently universal framework will serve as the fulcrum to facilitate numerous advanced nonlocal devices of enhanced performance and increased manufacturability, such as antenna radomes, flat lenses, compact optical systems, and optical analog computers.

\section{Generalized Huygens' condition}
\label{Sec:GHC}
\subsection{Formulation and scattering analysis}
\label{Subsub:MetaAtomFormulation}

\begin{figure*}
    \includegraphics[width=0.75\textwidth]{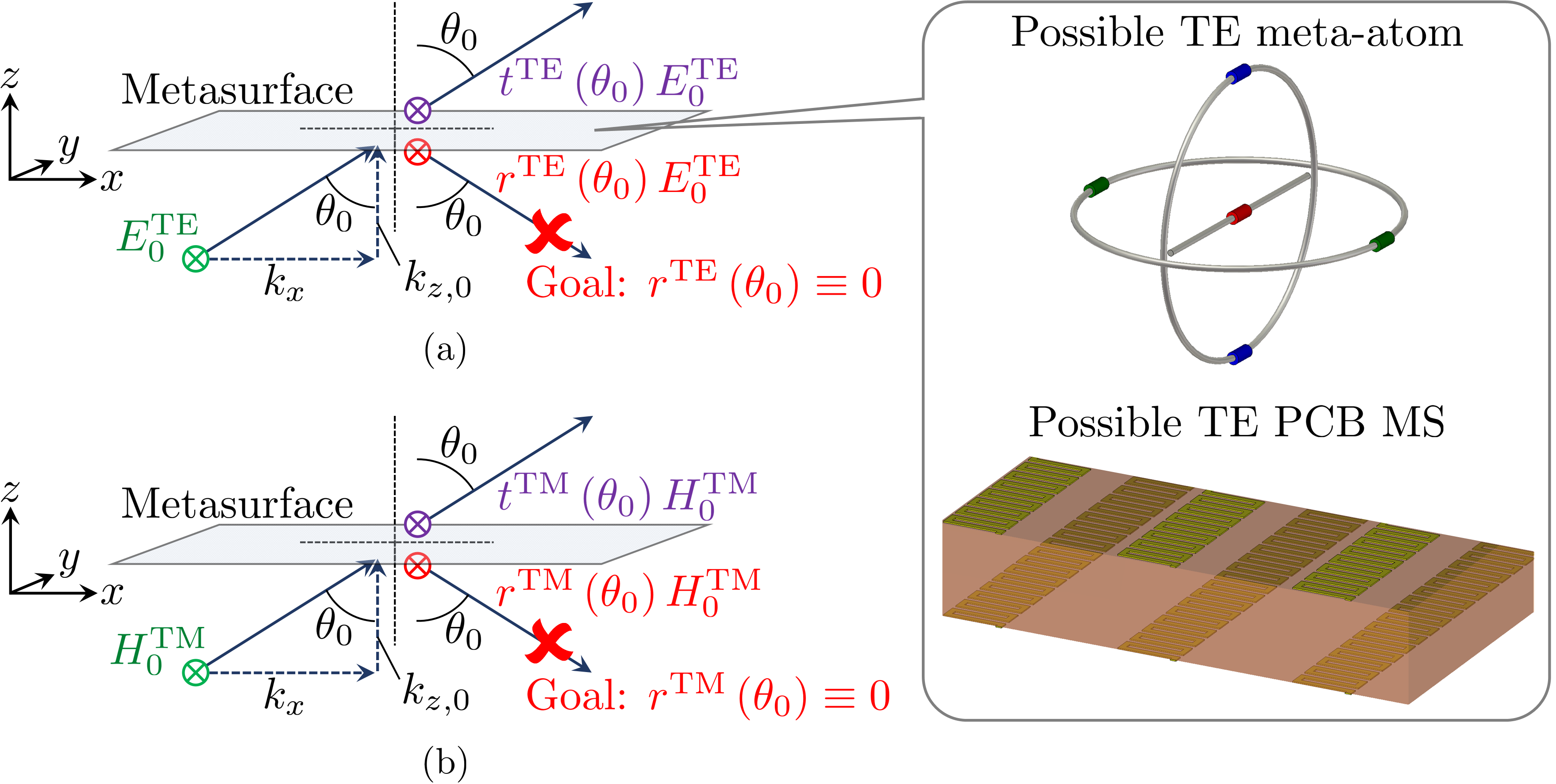}
\caption{Conceptual MS scattering setup decomposed into (a) TE- [Eq.\ (\ref{Eq:Ey})] and (b) TM- [Eq.\ (\ref{Eq:Hy})] polarized scenarios for the susceptibility configuration in Eq.\ (\ref{Eq:SusceptComponents}), along with the goal functionality of zero reflection at all angles of incidence. Wavevectors are represented by navy-blue arrows and decomposed into $x$ and $z$ wavenumbers (dashed arrows). The incident (green), reflected (red), and transmitted (purple) transverse field components [electric field in (a) and magnetic field in (b)] are represented by their values at the origin ($E_0^{\mathrm{TE}}$ and $H_0^{\mathrm{TM}}$). Inset: possible  meta-atom and MS level realizations (to be discussed throughout).}
\label{Fig:TETM}
\end{figure*}

Let us consider a homogenized MS lying on the $z=0$ plane in free space (Fig.\ \ref{Fig:TETM}), which, in general, can be modeled by the GSTCs \cite{Idemen1990,Tretyakov2003,Kuester2003,Achouri2015},
\begin{equation}
\label{Eq:GSTCs}
    \begin{aligned}
    \hat{z}\times\left( \vec{H}_{\mathrm{t}}^{+}-\vec{H}_{\mathrm{t}}^{-} \right)&=j\omega\vec{P}_{\mathrm{st}}-\hat{z}\times\vec{\nabla}_{\mathrm{t}}M_{\mathrm{s}z},\\
    \left( \vec{E}_{\mathrm{t}}^{+}-\vec{E}_{\mathrm{t}}^{-}\right)\times\hat{z}&=j\omega\mu_{0}\vec{M}_{\mathrm{st}}-\vec{\nabla}_{\mathrm{t}}\left( \frac{P_{\mathrm{s}z}}{\epsilon_{0}}\right)\times\hat{z},
    \end{aligned}
\end{equation}
and
\begin{equation}
\label{Eq:Suscept}
    \begin{aligned}
		\vec{P_{\mathrm{s}}}&=\epsilon_{0}\dbar{\chi}_{\mathrm{ee}}\cdot\vec{E}^{\mathrm{av}}+c^{-1}\dbar{\chi}_{\mathrm{em}}\cdot\vec{H}^{\mathrm{av}},\\	\vec{M}_{\mathrm{s}}&=\eta_{0}^{-1}\dbar{\chi}_{\mathrm{me}}\cdot\vec{E}^{\mathrm{av}}+\dbar{\chi}_{\mathrm{mm}}\cdot\vec{H}^{\mathrm{av}},
	\end{aligned}
\end{equation}
where $\pm$ superscripts denote the fields ($\vec{E}$ and $\vec{H}$) as evaluated at $z\to 0^{\pm}$; $\mathrm{t}$ and $z$ subscripts denote components tangential and normal to the MS, respectively; $\mathrm{av}$ superscript denotes the average of the fields acting on the MS, i.e., $\vec{E}^{\mathrm{av}}=\frac{1}{2}\left(\vec{E}^{+}+\vec{E}^{-}\right)$ and $\vec{H}^{\mathrm{av}}=\frac{1}{2}\left(\vec{H}^{+}+\vec{H}^{-}\right)$; $\vec{P}_{\mathrm{s}}$ and $\vec{M}_{\mathrm{s}}$ are the macroscopic distributions of surface electric and magnetic polarizations on the MS, respectively; $\dbar{\chi}_{\mathrm{ee}}$, $\dbar{\chi}_{\mathrm{mm}}$, $\dbar{\chi}_{\mathrm{em}}$, and $\dbar{\chi}_{\mathrm{me}}$ are the electric, magnetic, electro-magnetic, and magneto-electric susceptibility tensors, respectively; $\epsilon_{0}$, $\mu_{0}$, $\eta_{0}$, and $c$ are the standard constitutive parameters of free space; and harmonic time dependence of $e^{j\omega t}$ is assumed and suppressed. 

The susceptibility tensors ($\dbar{\chi}$) are directly determined by the materials and geometries comprising the MS. In its general form, each susceptibility tensor encompasses $3\times3$ dyadic components,
\begin{equation}
\label{Eq:Tensor}
    \dbar{\chi}=
    \begin{bmatrix}
        \chi^{xx}   &\chi^{xy}  &\chi^{xz}\\
        \chi^{yx}   &\chi^{yy}  &\chi^{zz}\\
        \chi^{zx}   &\chi^{zy}  &\chi^{zz}
    \end{bmatrix},
\end{equation}
which assess the impact of each field component ($E_x$, $E_y$, $E_z$, $H_x$, $H_y$, and $H_z$) on each polarization component ($P_{\mathrm{s}x}$, $P_{\mathrm{s}y}$, $P_{\mathrm{s}z}$, $M_{\mathrm{s}x}$, $M_{\mathrm{s}y}$, and $M_{\mathrm{s}z}$).
It is important to emphasize that the constitutive relations in Eq.\ (\ref{Eq:Suscept}) are described in an absolutely \emph{local} manner, i.e., each polarization component at a given location on the MS responds \emph{only} to the relevant field component at the exact same location (to a very good approximation). Hence, the meta-atom level spatially dispersive effects in this section are to be emulated by the inherent nonlocality of Maxwell's equations \emph{per se} (due to the spatial derivatives that relate the field components to one another).

As discussed in Sec.\ \ref{Sec:Intro}, our goal herein is to derive adequate relations between the various susceptibility components to simultaneously suppress Fresnel reflection for all angles of incidence. Expecting that such a functionality would necessitate additional extent of spatially dispersive properties, we purposely include both tangential and normal susceptibility components. The susceptibility tensors to be considered henceforth assume electric and magnetic anisotropy, both of whose principal axes are co-aligned with the $\left(x,y,z\right)$ coordinate system, i.e.,
\begin{equation}
\label{Eq:SusceptComponents}
    \begin{aligned}
    &\dbar{\chi}_{\mathrm{ee}}=
    \begin{bmatrix}
        \chi_{\mathrm{ee}}^{xx} &0  &0\\
        0   &\chi_{\mathrm{ee}}^{yy}    &0\\
        0   &0   &\chi_{\mathrm{ee}}^{zz}
    \end{bmatrix}
    ,
    &\dbar{\chi}_{\mathrm{mm}}=
    \begin{bmatrix}
        \chi_{\mathrm{mm}}^{xx} &0  &0\\
        0   &\chi_{\mathrm{mm}}^{yy}    &0\\
        0   &0   &\chi_{\mathrm{mm}}^{zz}
    \end{bmatrix},\\
    &\dbar{\chi}_{\mathrm{em}}=\dbar{\chi}_{\mathrm{me}}=0.
    \end{aligned}
\end{equation}
Such a type of configuration is reciprocal as it satisfies
\begin{equation}
 \label{Eq:Reciprocity}
    \dbar{\chi}_{\mathrm{ee}}^{\mathrm{T}}=\dbar{\chi}_{\mathrm{ee}},\; \dbar{\chi}_{\mathrm{mm}}^{\mathrm{T}}=\dbar{\chi}_{\mathrm{mm}},\; \dbar{\chi}_{\mathrm{me}}^{\mathrm{T}}=-\dbar{\chi}_{\mathrm{em}},
\end{equation}
where $\left(\cdot\right)^{\mathrm{T}}$ denotes matrix transposition \cite{Pozar2012,Achouri2015,Pfeiffer2016Emulating,Asadchy2020}. This is a necessary condition to avoid realization schemes that require complex measures to violate reciprocity, e.g., nonlinear, active, or time-modulated elements, or external magnetic bias \cite{Kodera2011,Pozar2012,Radi2016,Sounas2017,Sounas2018,Asadchy2020}. Assumed lossless as well, the susceptibility tensors satisfy
\begin{equation}
\label{Eq:Lossless}
    \dbar{\chi}_{\mathrm{ee}}^{\mathrm{T}}=\dbar{\chi}_{\mathrm{ee}}^{*},\; \dbar{\chi}_{\mathrm{mm}}^{\mathrm{T}}=\dbar{\chi}_{\mathrm{mm}}^{*},\; \dbar{\chi}_{\mathrm{me}}^{\mathrm{T}}=\dbar{\chi}_{\mathrm{em}}^{*},
\end{equation}
where $\left(\cdot\right)^{*}$ denotes elementwise complex conjugation \cite{Achouri2015}. Hence, for passive and lossless constituents, all the susceptibility components in Eq.\ (\ref{Eq:SusceptComponents}) are purely real valued.

As defined in Eq.\ (\ref{Eq:SusceptComponents}), we focus on macroscopically uniform configurations, for which the MS susceptibilities ($\dbar{\chi}$) are translationally invariant along the transverse coordinates, $x$ and $y$. Consequently, if the structure is excited by a plane wave of a certain transverse wavevector $\vec{k}_{\mathrm{t}}$, the scattered fields everywhere consist solely of its fundamental Floquet-Bloch (FB) harmonic (zeroth diffraction order); namely, the scattered waves everywhere correspond to the same tangential wavevector $\vec{k}_{\mathrm{t}}$ \cite{Tretyakov2003}. From now on, for the sake of simplicity, let us focus on the elementary incidence scenario for which the tangential wavevector $\vec{k}_{\mathrm{t}}$ coincides with one of the principal axes, say $x$, i.e., $\vec{k}_{\mathrm{t}}=k_{x}\hat{x}$. The corresponding normal wavenumber is given via $k_{z,0}=\sqrt{k_{0}^{2}-k_{x}^{2}}$, where $k_0=\omega/c$ is the wavenumber in free space.

Similarly to any plane-wave scattering problem off a homogeneous planar interface, this sets the $xz$-plane as the plane of incidence, such that each wave can be decomposed into a transverse-electric (TE) component, which involves only the $E_y$, $H_x$, and $H_z$ fields, and transverse-magnetic (TM) component, which involves only the $H_y$, $E_x$, and $E_z$ fields \cite{Jackson1999,Balanis2012,Pozar2012}. As such, any plane wave impinging the MS from below ($z<0$) at angle $\theta_0$ can be expressed via the $y$-directed transverse fields $E_{y}^{\mathrm{inc}}(\vec{r})=E_{0}^{\mathrm{TE}}e^{-j(k_{x}x+k_{z,0}z)}$ and $H_{y}^{\mathrm{inc}}(\vec{r})=H_{0}^{\mathrm{TM}}e^{-j(k_{x}x+k_{z,0}z)}$, where $k_x=k_0\sin\theta_0$ and $k_{z,0}=k_0\cos\theta_0$ are the wavenumbers tangential and normal to the MS, respectively, and
$E_{0}^{\mathrm{TE}}$ and $H_{0}^{\mathrm{TM}}$ are the transverse field amplitudes at the origin (Fig.\ \ref{Fig:TETM}).

Following the GSTCs, Eqs.\ (\ref{Eq:GSTCs}) and (\ref{Eq:Suscept}), along with the susceptibility configuration of Eq.\ (\ref{Eq:SusceptComponents}), we notice that the $\chi_{\mathrm{ee}}^{yy}$, $\chi_{\mathrm{mm}}^{xx}$, and $\chi_{\mathrm{mm}}^{zz}$ components interact \emph{only} with the TE field components ($E_{y}$, $H_{x}$, and $H_{z}$), whereas the $\chi_{\mathrm{mm}}^{yy}$, $\chi_{\mathrm{ee}}^{xx}$ and $\chi_{\mathrm{ee}}^{zz}$ interact \emph{only} with the TM components ($H_{y}$, $E_{x}$, and $E_{z}$). Therefore, the two polarizations are totally decoupled: TE-polarized illumination results solely in TE-polarized scattered waves, while TM-polarized illumination allows only TM-polarized scattered waves. In other words, TE-polarized incident waves do not excite any TM-polarized waves and vice versa.

According to the framework established above, the transverse fields everywhere can be expressed via
\begin{equation}
\label{Eq:Ey}
    \begin{aligned}
        E_y\left(\vec{r}\right)=&E_{0}^{\mathrm{TE}}e^{-jk_{x}x}\\
        &\times
        \begin{cases}
            e^{-jk_{z,0}z}+r^{\mathrm{TE}}\left(\theta_0\right)e^{+jk_{z,0}z}, &  z<0\\
            t^{\mathrm{TE}}\left(\theta_0\right)e^{-jk_{z,0}z}, & z>0
        \end{cases}\\
    \end{aligned}
\end{equation}
and
\begin{equation}
\label{Eq:Hy}
    \begin{aligned}
        H_y\left(\vec{r}\right)=&H_0^{\mathrm{TM}}e^{-jk_{x}x}\\
        &\times
        \begin{cases}
            e^{-jk_{z,0}z}+r^{\mathrm{TM}}\left(\theta_0\right)e^{+jk_{z,0}z}, &  z<0\\
            t^{\mathrm{TM}}\left(\theta_0\right)e^{-jk_{z,0}z}, & z>0,
        \end{cases}
    \end{aligned}
\end{equation}
where $r^{\mathrm{TE}}\left(\theta_0\right)$ and $t
^{\mathrm{TE}}\left(\theta_0\right)$ are the yet unknown field reflection and transmission coefficients corresponding to the TE specular reflection and direct transmission, respectively; similarly, $r^{\mathrm{TM}}\left(\theta_0\right)$ and $t
^{\mathrm{TM}}\left(\theta_0\right)$ are the scattering coefficients for the TM polarization.

To obtain the scattering coefficients defined in Eqs.\ (\ref{Eq:Ey}) and (\ref{Eq:Hy}) in terms of the MS constituents of Eq.\ (\ref{Eq:SusceptComponents}) and the angle of incidence $\theta_0$, we first substitute the fields of Eqs.\ (\ref{Eq:Ey}) and (\ref{Eq:Hy}) in Maxwell's equations to obtain expressions for the rest of the field components everywhere; next, we substitute the fields along with the susceptibilities of Eq.\ (\ref{Eq:SusceptComponents}) in the GSTCs of Eqs.\ (\ref{Eq:GSTCs}) and (\ref{Eq:Suscept}) and employ algebraic manipulations to extract the expressions for the scattering coefficients. The results to follow can be conveniently expressed once we normalize all the relevant quantities with respect to the wavenumber in free space $k_0$ to obtain unitless dimensions. The wavevector components are normalized according to
\begin{equation}
\label{Eq:kNormalization}
    \widetilde{k}_{x}=\frac{k_{x}}{k_0}=\sin\theta_0,
    \quad
    \widetilde{k}_{z,0}=\frac{k_{z,0}}{k_0}=\cos\theta_0,
\end{equation}
while the susceptibilities $\chi$ [which, as defined herein in Eq.\ (\ref{Eq:Suscept}), assume length dimensions] are normalized via
\begin{equation}
\label{Eq:ChiNormalization}
    \widetilde{\chi}=k_0\chi.
\end{equation}
Overall, for the TE part, this analysis yields
\begin{equation}
\label{Eq:scatTE}
    \begin{aligned}
    r^{\mathrm{TE}}\left(\theta_0\right)&=\frac{r^{\mathrm{TE}}_{0}+r^{\mathrm{TE}}_{2}\widetilde{k}_{z,0}^{2}}{d^{\mathrm{TE}}_{0}+d^{\mathrm{TE}}_{1}\widetilde{k}_{z,0}+d^{\mathrm{TE}}_{2}\widetilde{k}_{z,0}^{2}+d^{\mathrm{TE}}_{3}\widetilde{k}_{z,0}^{3}},\\
    t^{\mathrm{TE}}\left(\theta_0\right)&=\frac{t^{\mathrm{TE}}_{1}\widetilde{k}_{z,0}+t^{\mathrm{TE}}_{3}\widetilde{k}_{z,0}^{3}}{d^{\mathrm{TE}}_{0}+d^{\mathrm{TE}}_{1}\widetilde{k}_{z,0}+d^{\mathrm{TE}}_{2}\widetilde{k}_{z,0}^{2}+d^{\mathrm{TE}}_{3}\widetilde{k}_{z,0}^{3}},
    \end{aligned}
\end{equation}
where
\begin{equation}
\label{Eq:TECoeff}
    \begin{aligned}
        &\begin{aligned}
            &r^{\mathrm{TE}}_0=-2\left( \widetilde{\chi}_{\mathrm{ee}}^{yy}+\widetilde{\chi}_{\mathrm{mm}}^{zz}\right),
            &r^{\mathrm{TE}}_2=2\left( \widetilde{\chi}_{\mathrm{mm}}^{xx}+\widetilde{\chi}_{\mathrm{mm}}^{zz}\right),
        \end{aligned}\\
        &t^{\mathrm{TE}}_1=-j\left[\right(\widetilde{\chi}_{\mathrm{ee}}^{yy}+\widetilde{\chi}_{\mathrm{mm}}^{zz}\left) \widetilde{\chi}_{\mathrm{mm}}^{xx}+4\right],\\
        &\begin{aligned}
            &t^{\mathrm{TE}}_3=j \widetilde{\chi}_{\mathrm{mm}}^{xx}\widetilde{\chi}_{\mathrm{mm}}^{zz},
            &d^{\mathrm{TE}}_0=2\left( \widetilde{\chi}_{\mathrm{ee}}^{yy}+\widetilde{\chi}_{\mathrm{mm}}^{zz}\right),
        \end{aligned}\\
        &d^{\mathrm{TE}}_{1}=j\left[ \left( \widetilde{\chi}_{\mathrm{ee}}^{yy}+\widetilde{\chi}_{\mathrm{mm}}^{zz}\right)\widetilde{\chi}_{\mathrm{mm}}^{xx}-4\right],\\
        &\begin{aligned}
            &d^{\mathrm{TE}}_2=2\left( \widetilde{\chi}_{\mathrm{mm}}^{xx}-\widetilde{\chi}_{\mathrm{mm}}^{zz}\right),
            &d^{\mathrm{TE}}_3&=-j\widetilde{\chi}_{\mathrm{mm}}^{xx}\widetilde{\chi}_{\mathrm{mm}}^{zz}
        \end{aligned}
    \end{aligned}
\end{equation}
are the respective coefficients. The TM scattering coefficients take the same rational form of Eqs.\ (\ref{Eq:scatTE}) and (\ref{Eq:TECoeff}) subject to replacing each TE susceptibility component ($\widetilde{\chi}_{\mathrm{ee}}^{yy}$, $\widetilde{\chi}_{\mathrm{mm}}^{xx}$, and $\widetilde{\chi}_{\mathrm{mm}}^{zz}$) by its dual TM component ($\widetilde{\chi}_{\mathrm{mm}}^{yy}$, $\widetilde{\chi}_{\mathrm{ee}}^{xx}$, and $\widetilde{\chi}_{\mathrm{ee}}^{zz}$) as follows
\begin{equation}
\label{Eq:Substitutions}
    \begin{matrix}
        &\widetilde{\chi}_{\mathrm{ee}}^{yy}\to\widetilde{\chi}_{\mathrm{mm}}^{yy},
        &\widetilde{\chi}_{\mathrm{mm}}^{xx}\to\widetilde{\chi}_{\mathrm{ee}}^{xx},
        &\widetilde{\chi}_{\mathrm{mm}}^{zz}\to\widetilde{\chi}_{\mathrm{ee}}^{zz};
    \end{matrix}
\end{equation}
the explicit expressions are provided in Appendix \ref{App:TMexpressions}.

We observe that the scattering coefficients establish rational functions of $\widetilde{k}_{z,0}=\cos\theta_0$ (the $\widetilde{k}_{x}$-dependence is eliminated by the trigonometric identity $\widetilde{k}_{x}^{2}=1-\widetilde{k}_{z}^{2}$ in the derivation process), whose coefficients are exclusively determined by the susceptibility values embedded in the MS. One powerful way to interpret these results is that of angular filters \cite{Mailloux1976,Franchi1983,Ortiz2013,Shaham2022} (also known as spatial filters), in analogy to conventional circuit filters in frequency domain \cite{Matthaei1980,Winder2002,Pozar2012}. Herein, the normal wavenumber $\widetilde{k}_{z,0}$ in the expressions for the scattering coefficients plays the role of frequency $\omega$ in the rational transfer functions of circuit filters \cite{Shaham2022}. By tuning the susceptibility values of Eq.\ (\ref{Eq:SusceptComponents}), one may control the locations of the zeros and poles in the complex plane of the equivalent Laplace parameter $s=j\widetilde{k}_{z,0}$ (in analogy to the Laplace parameter in frequency domain $s=j\omega$) and thus to determine the type of the angular filter (low-pass, high-pass, etc.), to adjust its 3-dB cutoff angles, and so forth.

Hence, in comparison to other expressions of MS scattering coefficients available in the literature, e.g., \cite{Tretyakov2003,Holloway2005,Radi2015,Albooyeh2017,Momeni2021}, such an intuitive rational form of Eq.\ (\ref{Eq:scatTE}) that depends only on a single explicit angular parameter ($\widetilde{k}_{z,0}$) and on local constitutive properties of the MS (angularly independent susceptibility components $\chi$) is highly useful in the context of spatial dispersion. It unravels the distinct role of each component in the overall angular behavior and allows one to instantly draw simple yet powerful conclusions as to the functionality of the structure. Particularly, it charts a clear path towards all-pass angular characteristics and microscopic meta-atom implementations, as we shall soon demonstrate.

\subsection{All-angle transparency}
\label{Sub:AllAngle}
Recalling our purpose of omnidirectional transparency, we seek to eliminate reflection at all angles of incidence. In the context of angular-filter theory, this can be viewed as implementing an all-pass filter. To this end, we enforce the reflection coefficient to vanish for all angles of incidence, i.e., $r^{\mathrm{TE}}\left(\theta_0\right)\equiv 0$ and $r^{\mathrm{TM}}\left(\theta_0\right)\equiv 0$. For the TE polarization, this requires $r^{\mathrm{TE}}_{0}=r^{\mathrm{TE}}_{2}=0$, which yields the TE GHC [see Eq.\ (\ref{Eq:TECoeff})]
\begin{equation}
\label{Eq:GenHuygensTE}
    \widetilde{\chi}_{\mathrm{mm}}^{xx}=\widetilde{\chi}_{\mathrm{ee}}^{yy}=-\widetilde{\chi}_{\mathrm{mm}}^{zz}\triangleq\widetilde{\chi}_{\mathrm{GHC}}^{\mathrm{TE}};
\end{equation}
similarly, the dual GHC for the TM polarization follows Eq.\ (\ref{Eq:GenHuygensTE}) by replacing the TE components with their dual TM ones [see Eq.\ (\ref{Eq:Substitutions}) and Appendix \ref{App:TMexpressions}], i.e.,
\begin{equation}
\label{Eq:GenHuygensTM}
    \widetilde{\chi}_{\mathrm{ee}}^{xx}=\widetilde{\chi}_{\mathrm{mm}}^{yy}=-\widetilde{\chi}_{\mathrm{ee}}^{zz}\triangleq\widetilde{\chi}_{\mathrm{GHC}}^{\mathrm{TM}}.
\end{equation}
Here we have defined the common susceptibility values $\widetilde{\chi}_{\mathrm{GHC}}^{\mathrm{TE}}$ and $\widetilde{\chi}_{\mathrm{GHC}}^{\mathrm{TM}}$, which serve as the only degrees of freedom to control the angular profile of the transmission coefficients $t^{\mathrm{TE}}\left(\theta_0\right)$ and $t^{\mathrm{TM}}\left(\theta_0\right)$. For lossless MSs ($\widetilde{\chi}_{\mathrm{GHC}}^{\mathrm{TE}},\widetilde{\chi}_{\mathrm{GHC}}^{\mathrm{TM}}\in\mathbb{R}$), such control is exerted on the transmission phase, $\angle t^{\mathrm{TE}}\left(\theta_0\right)$ and $\angle t^{\mathrm{TM}}\left(\theta_0\right)$, as the transmission magnitude is identically unity (power conservation).

Focusing on the conditions
relevant to TE-polarized fields, let us further examine the GHC obtained in Eq.\ (\ref{Eq:GenHuygensTE}). The first necessary sub-condition is that of balanced \emph{tangential} electric and magnetic responses, $\widetilde{\chi}_{\mathrm{ee}}^{yy}=\widetilde{\chi}_{\mathrm{mm}}^{xx}$. In view of Eqs.\ (\ref{Eq:scatTE}) and (\ref{Eq:TECoeff}), this sub-condition is, in fact, the aforementioned standard Huygens' condition for normal incidence, $r^{\mathrm{TE}}\left(\theta_0=0\right)=0$ \cite{Pfeiffer2013Huygens,Monticone2013,Selvanayagam2013,Pfeiffer2013Cascaded,Pfeiffer2013Millimeter,Wong2014,Epstein2016,EpsteinNature2016}. However, Eq.\ (\ref{Eq:GenHuygensTE}) entails another sub-condition to be satisfied, that is, $\widetilde{\chi}_{\mathrm{ee}}^{yy}=-\widetilde{\chi}_{\mathrm{mm}}^{zz}$. Such a requirement implies separate distinct balance between each tangential response and its dual normal response in opposite sign. This is, in fact, Huygens' condition for vanishing reflection at \emph{grazing} incidence ($\theta_0\to90^{\circ}$), namely $r^{\mathrm{TE}}(\theta_0\to90^{\circ})=0$. It can be verified by taking the limit $\theta_0\to90^{\circ}$, i.e., $\widetilde{k}_{z,0}\to0$ in Eq.\ (\ref{Eq:scatTE}): unless the middle sub-condition of Eq.\ (\ref{Eq:GenHuygensTE}) is satisfied, the reflection coefficient for grazing angles of incidence inexorably approaches to $\frac{r^{\mathrm{TE}}_0}{d^{\mathrm{TE}}_0}=-1$ while the transmission coefficient wanes to $0$.

Indeed, it is naturally expected that the GHC, which eliminates backscattering at all angles of incidence, would specifically necessitate such elimination at the normal ($\theta_0=0$) and grazing ($\theta_0\to90^{\circ}$) angles. However, the converse, astonishingly, holds as well. Namely, to achieve all-angle transparency, it is sufficient to stipulate zero reflection solely at the two special angles of normal and grazing incidence, without any further requirements.

This phenomenon can be explained as follows. For the normal angle of incidence $\theta_0=0$, the normal field $H_z$ is absent, such that no normal polarization $M_{\mathrm{s}z}$ is induced. In this regime, the normal susceptibility $\widetilde{\chi}_{\mathrm{mm}}^{zz}$ is redundant and the scattering properties coincide with those of the existing tangentially polarizable Huygens' MSs \cite{Pfeiffer2013Huygens,Monticone2013,Selvanayagam2013,Pfeiffer2013Cascaded,Pfeiffer2013Millimeter,Wong2014,Epstein2016,EpsteinNature2016}. Hence, the Huygens' condition for normal incidence, $\chi_{\mathrm{ee}}^{yy}=\chi_{\mathrm{mm}}^{xx}$, which manifests balanced reflected fields that interfere destructively at this angle \cite{Love1976,Ziolkowski2010,Ziolkowski2022}, yet remains a necessary condition to achieve the more general requirement of all-angle vanishing reflection. However, for grazing incidence ($\theta_0\to90^{\circ}$), the tangential magnetic field vanishes ($H_{x}\to0$) and, thereby, the tangential magnetic polarization vanishes as well ($M_{\mathrm{s}x}\to0$). The \emph{only} possible way to balance such an absent magnetic polarization at grazing incidence is to ensure perfect cancellation of the remaining non-negligible tangential polarization ($P_{\mathrm{s}y}$) by harnessing the gradient of the normal magnetic one [$\hat{z}\times \vec{\nabla}_{\mathrm{t}}M_{\mathrm{s}z}$, Eq.\ (\ref{Eq:GSTCs})], such that the MS is rendered transparent at this angle [i.e., $t^{\mathrm{TE}}\left(\theta_0\to90^{\circ}\right)=1$].

Moreover, obliquely incident waves ($0<\theta_0<90^{\circ}$) introduce all the TE-polarized field components ($E_{y}$, $H_{x}$, and $H_{z}$). Consequently, when the GHC of Eq.\ (\ref{Eq:GenHuygensTE}) is satisfied, the ratio between the induced tangential magnetic current along $x$ ($j\omega\mu_0 M_{\mathrm{s}x}$) and the total effective $y$-directed electric current ($j\omega P_{\mathrm{s}y}-\partial _{x}M_{\mathrm{s}z}$) reads $-\eta_0^{-1}\widetilde{k}_{z,0}=-\eta_0^{-1}\cos\theta_0$ [according to Maxwell's equations and Eqs.\ (\ref{Eq:GSTCs}), (\ref{Eq:Suscept}), (\ref{Eq:Ey}), (\ref{Eq:kNormalization})--(\ref{Eq:TECoeff}), and (\ref{Eq:GenHuygensTE})]. This ratio coincides precisely with the negative of the TE-polarized wave admittance. From another perspective, the amplitude of the total magnetic polarization vector (composed of its $x$ and $z$ projections) for any oblique angle $\theta_0$ is balanced with respect to the electric polarization (ratio of $\eta_0^{-1}$), and its direction is perpendicular to both the electric polarization and the wavevector of the absent specularly reflected wave. According to both these equivalent interpretations, the electric and magnetic polarizations balance each other in the sense that the relevant projections of backscattered fields emanated off the former cancel those off the latter for all angles. Thus, for oblique, and particularly grazing, angles, the normal susceptibility component plays a substantial part in maintaining balance and suppressing reflection. 

Overall, we conclude that the GHC required to repress reflection for all angles of incidence is equivalent to combining two distinct fundamental Huygens' conditions: the one for normal incidence and that for grazing incidence. Similar corollaries can be achieved for the dual TM-polarized fields as well by following the replacements prescribed in Eq.\ (\ref{Eq:Substitutions}). Importantly, we pay attention to the fragile binary nature of scattering at the extreme regime of grazing incidence: deviation from the grazing-angle Huygens' condition, albeit small, inevitably leads to unity reflectance and vanishing transmittance at this angle. Such a sharp sensitivity indicates that Huygens' condition for grazing incidence should be top prioritized in all-angle reflectionless designs.

\begin{figure}
    \includegraphics[width=0.42\textwidth]{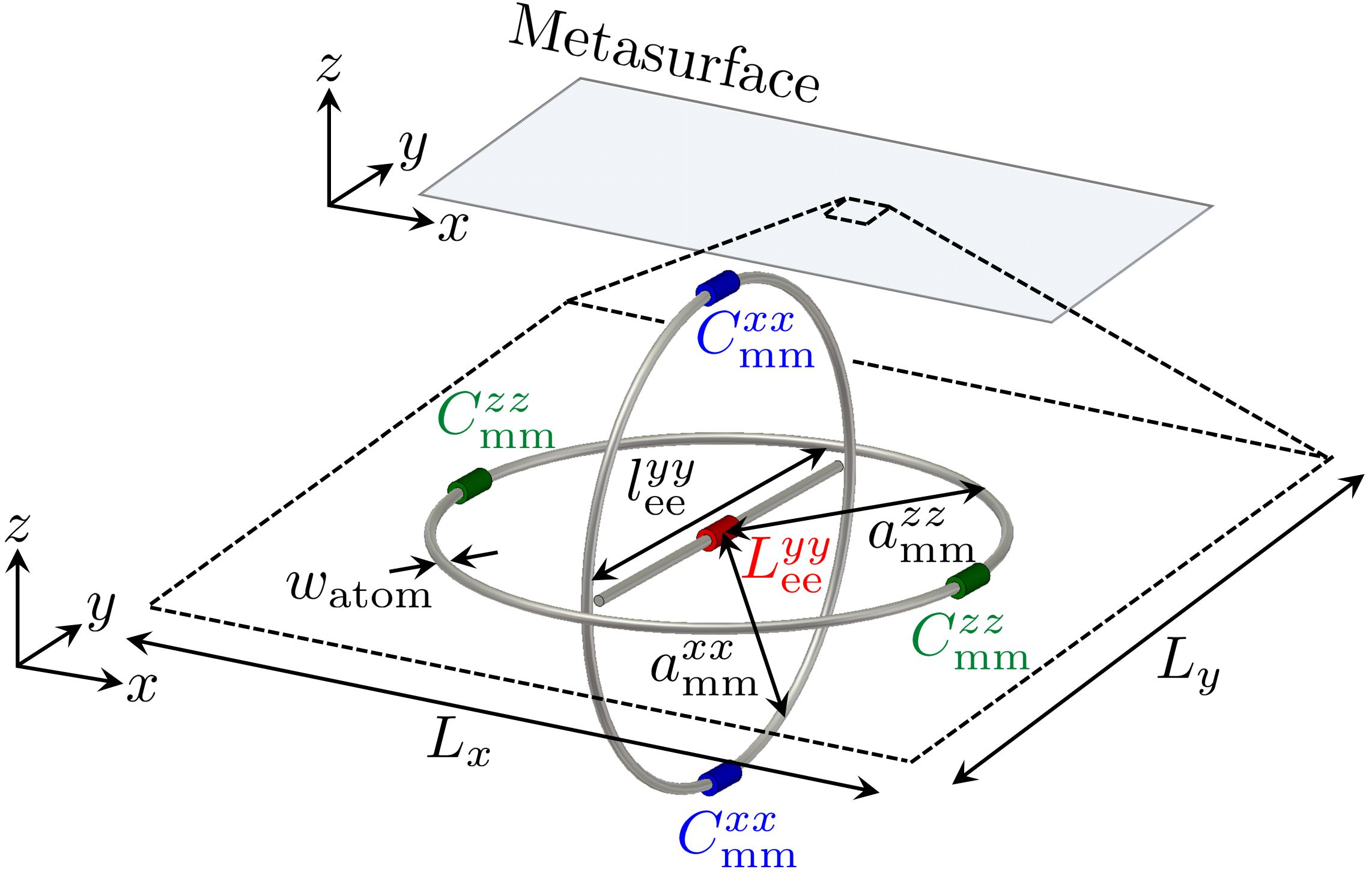}
    \caption{Physical meta-atom configuration for realizing the TE GHC of Eq.\ (\ref{Eq:GenHuygensTE}) at $f=20$ GHz. $L_{x}=L_{y}=3$ mm $\approx0.2\lambda_0$ are the unit-cell sizes along $x$ and $y$, respectively; $w_{\mathrm{atom}}=0.03$ mm $\approx0.002\lambda_0$ is the wire diameter common to all inclusions; $a_{\mathrm{mm}}^{zz}=0.75$ mm $\approx$ $0.05 \lambda_0$ and $a_{\mathrm{mm}}^{xx}=0.675$ mm $\approx 0.045\lambda_0$ are the outer radii of the in-plane ($\widetilde{\chi}_{\mathrm{mm}}^{zz}$) and reclined ($\widetilde{\chi}_{\mathrm{mm}}^{xx}$) capacitively loaded PEC loops; $l_{\mathrm{ee}}^{yy}=2\times 0.6$ mm $\approx 2\times 0.04\lambda_0$ is the total length of the straight inductively loaded PEC wire ($\widetilde{\chi}_{\mathrm{ee}}^{yy}$). Two lumped capacitive loads (small green cylinders) of $C_{\mathrm{mm}}^{zz}$ capacitance each are placed in-series to the in-plane loop; likewise, two capacitive loads ($C_{\mathrm{mm}}^{xx}$, blue cylinders) and one inductive load ($L_{\mathrm{ee}}^{yy}$ inductance, red cylinder) are placed on the reclined loop and the straight wire, respectively. Designed load values appear in Table \ref{Tab:FinalVals}.
    }
\label{Fig:MetaAtom}
\end{figure}

\begin{figure}
    \includegraphics[width=0.4\textwidth]{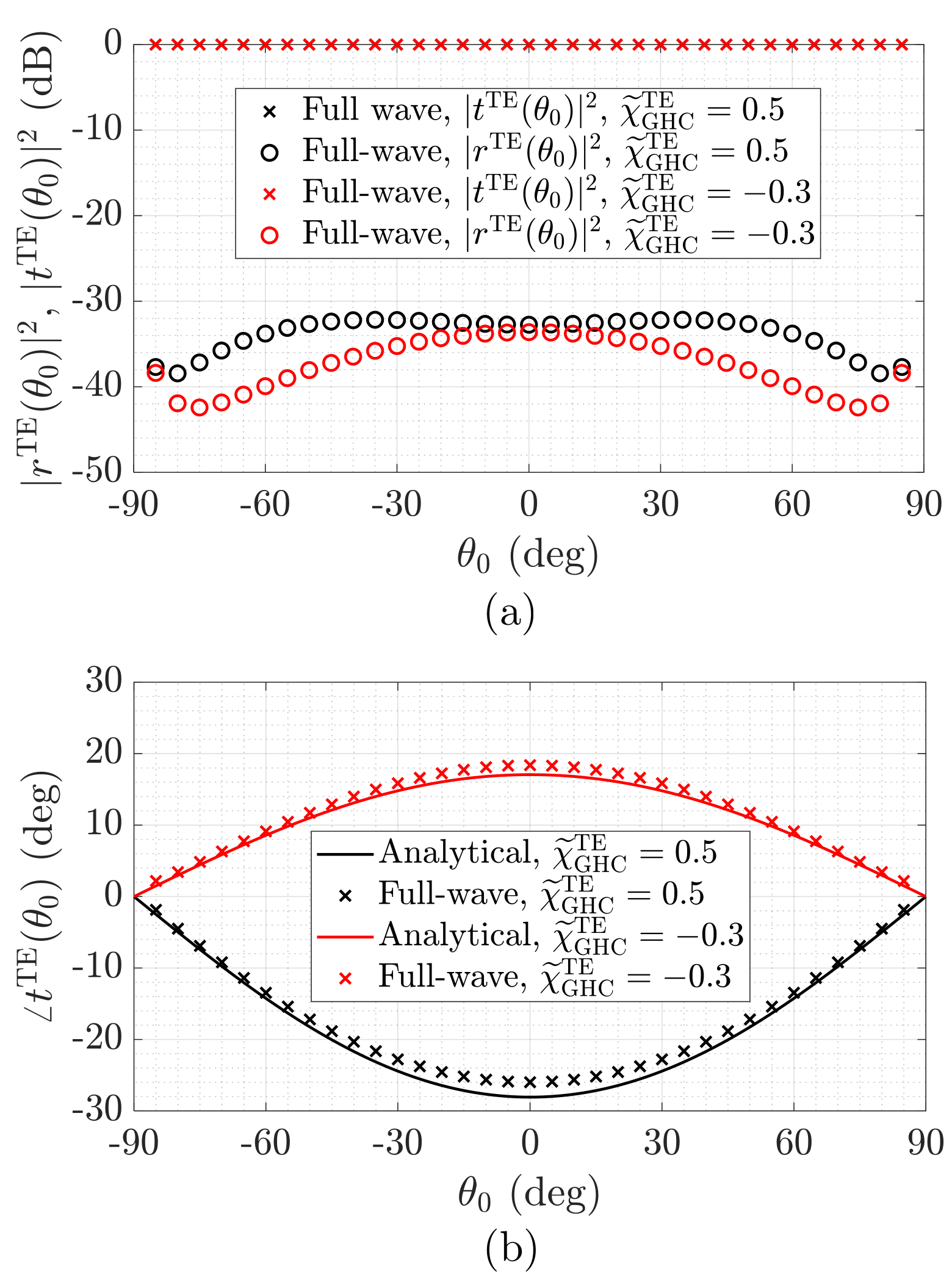}
\caption{Scattering results for the $\widetilde{\chi}_{\mathrm{GHC}}^{\mathrm{TE}}=0.5$ (black) and $\widetilde{\chi}_{\mathrm{GHC}}^{\mathrm{TE}}=-0.3$ (red) designs. (a) Full-wave reflection ($\circ$ markers) and transmission ($\times$ markers) magnitudes (dB) vs.\ angle of incidence. (b) Analytical (solid lines) and full-wave ($\times$ markers) results for the transmission phase vs.\ angle of incidence for both designs.}
\label{Fig:MAResults}
\end{figure}

\section{Omnidirectional transparency at the meta-atom level}
\label{Sec:MetaAtom}
To validate our formulation above, our aim now is to devise a physical structure capable of realizing the GHC at the meta-atom level. The demonstration herein shall be carried out at microwave frequencies. Nevertheless, we stress that the GHC is as universal as the classical Huygens' principle: it applies to all frequency ranges (e.g.,  optical wavelengths) and diverse physical wave phenomena \cite{Love1901,Pao1976,Xie2013,Monticone2013,Decker2015,Ye2016Making,Lekner2016,DiazRubio2017}.

We note that the surface-susceptibility-based GHC indicates a bottom-up implementation scheme to construct adequate meta-atom candidates. For the TE-polarized scenario, we are instructed to simultaneously endow the meta-atom with three kinds of response: tangential electric ($\widetilde{\chi}_{\mathrm{ee}}^{yy}$), tangential magnetic ($\widetilde{\chi}_{\mathrm{mm}}^{xx}$), and normal magnetic ($\widetilde{\chi}_{\mathrm{mm}}^{zz}$). Inspired by customary meta-atom synthesis techniques \cite{Tretyakov2003,Pfeiffer2013Huygens,Selvanayagam2013,Shaham2021,Shaham2022}, we propose the extended meta-atom geometry depicted in Fig.\ \ref{Fig:MetaAtom}: both the components of the magnetic response are realized via two perpendicular capacitively loaded loops of perfect-electric-conducting (PEC) wire; the tangential electric response is implemented via straight inductively loaded PEC wire concentric to the two loops.

Designating the frequency of operation to $f=\frac{\omega}{2\pi}=20$ GHz, we first set the unit-cell period to a deep-subwavelength size (at the behest of the homogenization approximation), $L_{x}=L_{y}=3$ mm $\approx0.2\lambda_0$ ($\lambda_0\approx15$ mm is free-space wavelength at $20$ GHz). Accordingly, we fix all the geometrical dimensions (detailed in the caption of Fig.\ \ref{Fig:MetaAtom}) so as to facilitate tunability within a wide assortment of susceptibility values \cite{Shaham2021}. Next, individually for each of the loops, we establish an accurate look-up table (LUT) that associates the physical capacitance values $C_{\mathrm{mm}}^{xx(zz)}$ to corresponding susceptibility values $\widetilde{\chi}_{\mathrm{mm}}^{xx(zz)}$, respectively (Appendix \ref{App:Char}). We achieve this by utilizing the characterization method outlined in our previous work \cite{Shaham2021}, which relies on the full-wave scheme in \cite{Zaluski2016} fitted with quasistatic analytical approximations \cite{Tretyakov2003}; the full-wave simulations throughout are performed in the commercial solver ``CST Microwave Studio'' (CST).

Given a desired goal value of $\widetilde{\chi}_{\mathrm{GHC}}^{\mathrm{TE}}$, we set the capacitance values $C_{\mathrm{mm}}^{xx(zz)}$, according to the LUT, to implement $\widetilde{\chi}_{\mathrm{mm}}^{xx}=\widetilde{\chi}_{\mathrm{GHC}}^{\mathrm{TE}}$ and $\widetilde{\chi}_{\mathrm{mm}}^{zz}=-\widetilde{\chi}_{\mathrm{GHC}}^{\mathrm{TE}}$, as instructed by the GHC of Eq.\ (\ref{Eq:GenHuygensTE}). Subsequently, to set the value of the load inductance $L_{\mathrm{ee}}^{yy}$, we first recall that the dominant requirement of the GHC is vanishing reflection at grazing angles $\theta_0\to90^{\circ}$. Therefore, we illuminate the meta-atom of Fig.\ \ref{Fig:MetaAtom} in CST by a plane wave at the near-grazing angle $\theta_0=85^{\circ}$ and seek for the specific $L_{\mathrm{ee}}^{yy}$ value that results in such vanishing reflectance, to thus guarantee $\widetilde{\chi}_{\mathrm{ee}}^{yy}=-\widetilde{\chi}_{\mathrm{mm}}^{zz}=\widetilde{\chi}_{\mathrm{GHC}}^{\mathrm{TE}}$. Due to the highly symmetrical geometry of the meta-atom, the circumventing quasistatic magnetic field emanated by the wire generates zero magnetic flux through the loops. As such, the magnetic coupling between the wire and the loops vanishes, such that the previously designed magnetic susceptibility values $\widetilde{\chi}_{\mathrm{mm}}^{xx}=\widetilde{\chi}_{\mathrm{GHC}}^{\mathrm{TE}}$ and $\widetilde{\chi}_{\mathrm{mm}}^{zz}=-\widetilde{\chi}_{\mathrm{GHC}}^{\mathrm{TE}}$ approximately remain intact regardless of the $L_{\mathrm{ee}}^{yy}$ value \cite{Wong2014,EpsteinNature2016}. This concludes the overall design procedure and sets all the susceptibility values according to the GHC.

To demonstrate and validate our theory and realization methodology, we follow the scheme above to design generalized Huygens' MSs for the TE polarization with typical susceptibility values of $\widetilde{\chi}_{\mathrm{GHC}}^{\mathrm{TE}}=0.5$ and $\widetilde{\chi}_{\mathrm{GHC}}^{\mathrm{TE}}=-0.3$; detailed steps and final load parameters are shown in Appendix \ref{App:Char} and Table \ref{Tab:FinalVals}. Next, we probe the performance of each MS in CST. Figure \ref{Fig:MAResults}(a) shows the reflectance $|r^{\mathrm{TE}}\left(\theta_0\right)|^{2}$ ($\circ$ markers) and transmittance $|t^{\mathrm{TE}}\left(\theta_0\right)|^{2}$ ($\times$ markers) for the $\widetilde{\chi}_{\mathrm{GHC}}^{\mathrm{TE}}=0.5$ (black) and $\widetilde{\chi}_{\mathrm{GHC}}^{\mathrm{TE}}=-0.3$ (red) case-study values vs.\ the angle of incidence at the range $-85^{\circ}\leq\theta_0\leq85^{\circ}$. Indeed, the reflectance practically vanishes for both the MSs ($|r^{\mathrm{TE}}\left(\theta_0\right)|^{2}<-32.17$ dB $\approx0.06\%$) as their transmittance essentially reaches unity ($|t^{\mathrm{TE}}\left(\theta_0\right)|^{2}>99.94\%$) for all the range of angles $\theta_0$. In addition, Fig.\ \ref{Fig:MAResults}(b) compares between the analytical predictions (solid lines) and full-wave results ($\times$ markers) of the transmission phase $\angle t^{\mathrm{TE}}\left(\theta_0\right)$ for both designs (black and red). Indeed, excellent agreement is observed between the theoretical and full-wave results, thus verifying the GHC theory presented above.

To conclude, we have thus far established the GHC and validated it at the meta-atom level. We have shown that the previously overlooked condition of vanishing reflection at grazing incidence, which has been found in the former section to play the dominant role in the all-angle response, indeed serves as a central demand to be manifested, even as an integral part of the design procedure itself. As mentioned earlier, we have, in fact, channeled the underlying meta-atom nonlocal mechanism of Maxwell's equations by balancing the contribution of each field component to the overall MS response. A generalization of the meta-atom concept in Fig.\ \ref{Fig:MetaAtom} that supports both TE and TM polarizations by incorporating also the latter part of the GHC [Eq.\ (\ref{Eq:GenHuygensTM})] in the design is demonstrated in Appendix \ref{App:AllPolarizations}; notably, this polarization insensitive implementation further retains omnidirectional transmittance for all other planes of incidence, in addition to the $xz$-plane considered in the original design (Appendix \ref{App:AllPolarizations}).

\section{Omnidirectional transparency at the metasurface level}
\label{Sec:MSLevel}
Having established the profound concept of the GHC, we now possess the means to utilize its results in more practical situations. Insightful as they are, the meta-atoms proposed in Fig.\ \ref{Fig:MetaAtom} may introduce fabrication challenges, as they are suspended in air. Indeed, a few previous reports proposed realization schemes to inhabit and align wire inclusions of somewhat similar nature, e.g., via low-permittivity plastic foam supports \cite{Asadchy2015PRX}. However, we aim at more rigid realizations that can serve, for example, as mechanically protective (yet electromagnetically transparent) radomes. Therefore, we resort to the standard and common practice to implement MSs in general, that is, planar multilayered configurations compatible with standard fabrication techniques \cite{Glybovski2016,Chen2016}.

In the recent past, such cascades of electrically polarizable sheets have proven versatile capability of realizing equivalent collocated electric, magnetic, and even magnetoelectric (chiral or omega type) in-plane susceptibilities \cite{Zhao2012Twisted,Pfeiffer2013Cascaded,Pfeiffer2013Millimeter,Pfeiffer2014Bian,Wong2014,Glybovski2016,Chen2016,Epstein2016OBMS,Epstein2016,Epstein2016PRL,Chen2018,Asadchy2018}. These designs typically tailor the excitation of symmetric and antisymmetric modes supported by the multilayered structure to emulate such meta-atom functionalities in practice. Specifically, at microwave frequencies, PCB-compatible designs of such constructs have proven themselves greatly appealing, as they are supported by standard, low-cost, and widely available fabrication techniques \cite{Pfeiffer2013Millimeter,Pfeiffer2014Bian,Epstein2016OBMS,Chen2018}.

\begin{figure*}
    \includegraphics[width=0.75\textwidth]{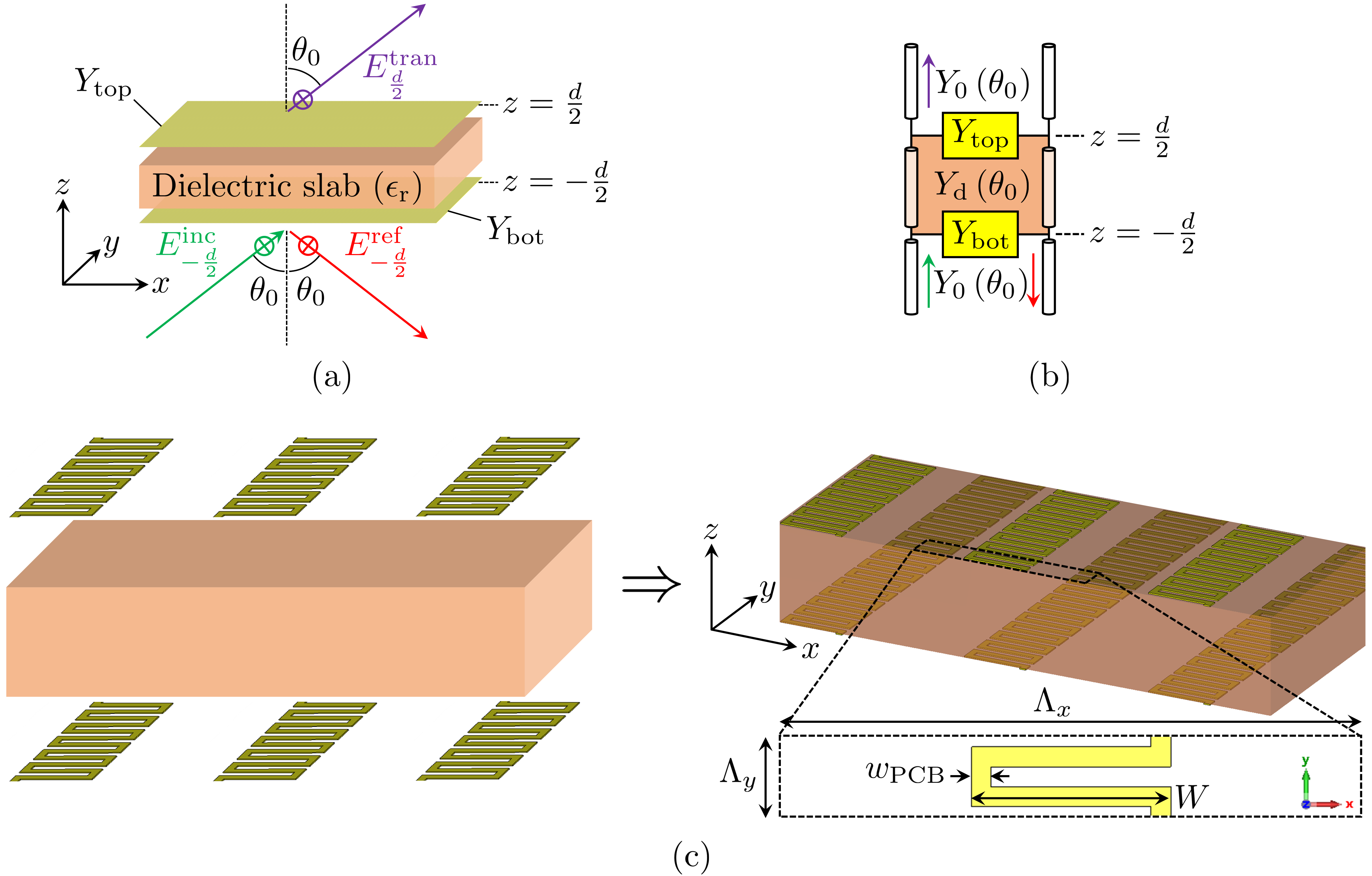}
\caption{(a) Physical scattering configuration for the PCB cascade: a dielectric substrate coated by two uniform admittance sheets, $Y_{\mathrm{bot}}$ and $Y_{\mathrm{top}}$; incident (green), reflected (red), and transmitted (purple) TE plane waves represented by transverse field values at the reference planes $z=-\frac{d}{2}$ and $z=\frac{d}{2}$. (b) Equivalent TL model for the scattering setup in (a): each region ($|z|>\frac{d}{2}$ and $|z|<\frac{d}{2}$) is modeled by a TL of respective propagation constant and characteristic admittance; the admittance sheets are modeled by shunt loads (yellow rectangles). (c) Copper meander-line realization of the admittance-sheet coatings in (a) for $f=20$ GHz and a commercial Rogers RO3003 substrate of $\epsilon_{\mathrm{r}}=3$ and $d=60$ mil. Inset: one unit-cell of periodicity $\Lambda_{x}=3$ mm $\approx0.2\lambda_0$ along $x$ and $\Lambda_{y}=4w_{\mathrm{PCB}}$ along $y$, where $w_{\mathrm{PCB}}=4$ mil is the trace width and standard $18\mu$m copper deposition thickness (0.5 oz.) is taken; the meander width $W$ is tuned to control the associated admittance value $Y_{\mathrm{top}}$ (or $Y_{\mathrm{bot}}$).}
\label{Fig:PCBConfig}
\end{figure*}

\subsection{Formulation and scattering analysis}
\label{Subsec:MSLevelFormulation}
We, hence, consider the simplest PCB-compatible configuration fathomable [in terms of analysis and standard fabrication complexity, Fig.\ \ref{Fig:PCBConfig}(a)], in which two admittance sheets, $Y_{\mathrm{bot}}$ and $Y_{\mathrm{top}}$, constitute the bottom ($z=-\frac{d}{2}$) and top ($z=\frac{d}{2}$) layers of the structure. These macroscopic (homogenized) admittance values are defined by \emph{locally} relating the electric field at each location of the layer to the induced tangential magnetic field discontinuity (surface current) therein,
\begin{equation}
\label{Eq:Ydefinition}
    \begin{aligned}
        \left({H_{x}^{+}-H_{x}^{-}}\right)_{z=-\frac{d}{2}}&=Y_{\mathrm{bot}}E_{y}(z=-\frac{d}{2}),\\
        \left({H_{x}^{+}-H_{x}^{-}}\right)_{z=\frac{d}{2}}&=Y_{\mathrm{top}}E_{y}(z=\frac{d}{2}).
    \end{aligned}
\end{equation}
A standard dielectric substrate of relative permittivity $\epsilon_{\mathrm{r}}>1$ is placed as mechanical support between these sheets ($-\frac{d}{2}<z<\frac{d}{2}$), such that the overall size of the slab is $d$. Recalling our cause of rendering this elementary cascade omnidirectionally transparent (yet mechanically rigid), we first set out analyze its scattering properties.

Once again, we illuminate the composite from below ($z<-\frac{d}{2}$) with a TE-polarized plane wave described via $E_{y}^{\mathrm{inc}}(\vec{r})=E_{0}e^{-j\left( k_{x}x+k_{z,0}z\right)}$; the angle of incidence $\theta_0$ remains related to the transverse, $k_{x}=k_0\sin\theta_0$, and normal, $k_{z,0}=k_0\cos\theta_0$, wavenumbers in free space; the normalized values, $\widetilde{k}_{x}$ and $\widetilde{k}_{z,0}$, are still defined via Eq.\ (\ref{Eq:kNormalization}). As before, the structure in question [Fig.\ \ref{Fig:PCBConfig}(a)] is macroscopically uniform along the transverse coordinates ($x$ and $y$). Therefore, only the fundamental FB harmonic (0th diffraction order) arises in the spectrum of the scattered fields. Consequently, all the scattered waves everywhere inherit the tangential wavenumber $k_{x}$ of the incident wave. In other words, only the specular reflection and direct transmission of normal wavenumbers ($\pm$)$\widetilde{k}_{z,0}$ appear in the free space regions, while refracted waves of normal wavenumber ($\pm$)$k_{z,\mathrm{d}}$, or, in its normalized form,
\begin{equation}
\label{Eq:kzd}
    \widetilde{k}_{z,\mathrm{d}}=\frac{k_{z,\mathrm{d}}}{k_0}=\sqrt{\epsilon_{\mathrm{r}}-\widetilde{k}_{x}^{2}}=\sqrt{\chi_{\mathrm{r}}+\widetilde{k}_{z,0}^2},
\end{equation}
appear in the dielectric; $\chi_{\mathrm{r}}=\epsilon_{\mathrm{r}}-1>0$ is the volumetric electrical susceptibility of the dielectric substrate (the wavenumbers $k_{x}$ and $k_{z,\mathrm{d}}$ are compatible with the classical Snell's law of refraction from free space into the dielectric and vice versa); the $+$ or $-$ signs preceding $k_{z,0}$ and $k_{z,\mathrm{d}}$ slightly before Eq.\ (\ref{Eq:kzd}) refer to forward and backward propagating waves, respectively.

In the free-space regions ($|z|>\frac{d}{2}$), which, for our purpose of MS scattering analysis, constitute the domain of interest, we may express the TE field as
\begin{equation}
\label{Eq:Ey_d}
    \begin{aligned}
        &E_y\left(\vec{r}\right)=e^{-jk_{x}x}\\
        &\times
        \begin{cases}
            E_{-\frac{d}{2}}^{\mathrm{inc}}e^{-jk_{z,0}\left(z+\frac{d}{2}\right)}+E_{-\frac{d}{2}}^{\mathrm{ref}}e^{+jk_{z,0}\left(z+\frac{d}{2}\right)}, &  z<-\frac{d}{2}\\
            E_{\frac{d}{2}}^{\mathrm{tran}}e^{-jk_{z,0}\left(z-\frac{d}{2}\right)}, & z>\frac{d}{2},
        \end{cases}
    \end{aligned}
\end{equation}
where $E_{-\frac{d}{2}}^{\mathrm{inc}}$ and $E_{-\frac{d}{2}}^{\mathrm{ref}}$ are the field amplitudes of the incident and reflected waves at the reference plane $z=-\frac{d}{2}$, and $E_{\frac{d}{2}}^{\mathrm{tran}}$ is the field amplitude of the transmitted wave at the reference plane $z=\frac{d}{2}$ \footnote{Note that Eq.\ (\ref{Eq:Ey_d}) is defined in a format slightly different from that of Eq.\ (\ref{Eq:Ey}), albeit both essentially describe similar scattered-field configurations of specular reflection and direct transmission. The reason is that care must be taken with respect to the finite MS thickness appertaining to Eq.\ (\ref{Eq:Ey_d}), compared to the zero MS thickness relevant to Eq.\ (\ref{Eq:Ey})}.

It also follows that the electromagnetic problem to be solved, in order to retrieve the relevant scattering coefficients, can be represented via equivalent TL model [Fig.\ \ref{Fig:PCBConfig}(b)] \cite{Pozar2012,Monticone2013,Pfeiffer2013Cascaded,Pfeiffer2013Millimeter,Epstein2016,Epstein2016OBMS}: propagation along the $z$ direction in free space (dielectric) is equivalent to propagation in a TL of \emph{angularly dependent} propagation constant $k_{z,0}$ ($k_{z,\mathrm{d}}$) and \emph{angularly dependent} characteristic admittance $Y_{0}=\eta_0^{-1}\widetilde{k}_{z,0}$ ($Y_{\mathrm{d}}=\eta_{0}^{-1}\widetilde{k}_{z,\mathrm{d}}$); according to Eq.\ (\ref{Eq:Ydefinition}), the bottom and top layers are equivalent to shunt loads of $Y_{\mathrm{bot}}$ and $Y_{\mathrm{top}}$ admittance values, respectively.

We follow the standard TL theory \cite{Pozar2012,Balanis2012} to analytically calculate the reflection $r_{-\frac{d}{2}\to -\frac{d}{2}}\left(\theta_0\right)$ and transmission $t_{-\frac{d}{2}\to \frac{d}{2}}\left(\theta_0\right)$ coefficients,
\begin{widetext}
\begin{equation}
    \label{Eq:RT_Cascade}
    \begin{split}
    r_{-\frac{d}{2}\to -\frac{d}{2}}\left(\theta_0\right)=
    \frac{E^{\mathrm{ref}}_{-\frac{d}{2}}}{E^{\mathrm{inc}}_{-\frac{d}{2}}}
    &=\frac{j\left[Y_0^2-Y_{\mathrm{d}}^2+Y_0\left(Y_{\mathrm{top}}-Y_{\mathrm{bot}}\right)-Y_{\mathrm{top}}Y_{\mathrm{bot}}\right]\tan \left(k_{z,\mathrm{d}}d\right)-Y_{\mathrm{d}}\left(Y_{\mathrm{top}}+Y_{\mathrm{bot}}\right)}
    {j\left[Y_0^2+Y_{\mathrm{d}}^2+Y_0\left(Y_{\mathrm{top}}+Y_{\mathrm{bot}}\right)+Y_{\mathrm{top}}Y_{\mathrm{bot}}\right]\tan \left(k_{z,\mathrm{d}}d\right)+Y_{\mathrm{d}}\left(2Y_0+Y_{\mathrm{top}}+Y_{\mathrm{bot}}\right)} \\
    t_{-\frac{d}{2}\to \frac{d}{2}}\left(\theta_0\right)=
    \frac{E^{\mathrm{tran}}_{\frac{d}{2}}}{E^{\mathrm{inc}}_{-\frac{d}{2}}}
    &=\frac{Y_{\mathrm{d}}\left[1+r_{-\frac{d}{2}\to -\frac{d}{2}}\left(\theta_0\right)\right]}{Y_{\mathrm{d}}\cos\left(k_{z,\mathrm{d}}d\right)+j(Y_0+Y_{\mathrm{top}})\sin\left(k_{z,\mathrm{d}}d\right)}.
    \end{split}
\end{equation}
\end{widetext}

It is evident, by the $\tan\left(k_{z,\mathrm{d}}d\right)$, $\cos\left(k_{z,\mathrm{d}}d\right)$, and $\sin\left(k_{z,\mathrm{d}}d\right)$ terms in Eq.\ (\ref{Eq:RT_Cascade}), that an additional type of nonlocal phenomenon, which did not appear in the meta-atom level design of Sec.\ \ref{Sec:MetaAtom}, takes place in the MS-level structure of Fig.\ \ref{Fig:PCBConfig}, that is, phase accumulation due to propagation across the slab ($e^{\mp jk_{z,\mathrm{d}}d}$). Such \emph{angularly dependent} phase delay originates from both refraction and multiple-reflection mechanisms entailed in the slab, which transversely convey echos of the incident wave along the MS \cite{Shastri2023}. Therefore, the ``input'' (incident) fields at a certain location on the MS influence the ``output'' (scattered) fields at different locations; equivalently, from the qualitative ray-tracing viewpoint, phase fronts incoming from different angles travel along different trajectories throughout the MS and hence result in scattered waves with different properties.

Furthermore, the MS-level design herein relies on yet another nonlocal mechanism of the meta-atom type, as encountered in Sec.\ \ref{Sec:MetaAtom}, which is inherited from the spatially dispersive Maxwell relations between the electric and magnetic fields in Eq.\ (\ref{Eq:Ydefinition}) (namely, $H_{x}=\frac{1}{j\omega\mu_0}\partial_{z}E_{y}$). By itself, each layer adhering Eq.\ (\ref{Eq:Ydefinition}) is merely electrically polarizable, and, therefore, is limited in terms of spatial dispersion, as it lacks the other magnetic components. Nevertheless, the MS nonlocal mechanism of phase accumulation shall aid the meta-atom one to enrich the total spatial dispersion of the scattering. As such, assuming a given dielectric substrate and a given frequency (i.e., given values of $\epsilon_{\mathrm{r}}$ and $k_{0}d$), we may tune the surface admittance values $Y_{\mathrm{top}}$ and $Y_{\mathrm{bot}}$ to shape the angular response of the scattering to best fit our liking.

\subsection{All-angle transparency}
\label{Subsec:CascadeGHC}
Bearing in mind, from our experience in Sec.\ \ref{Sec:MetaAtom}, that the grazing angle, $\widetilde{k}_{z,0}\to 0$ ($\theta_0\to 90^{\circ}$), is the hub around which all-angle transparency must be devised, we prioritize our transparency requirements to hold at such a regime. Specifically, aspiring to mimic free-space propagation along distance $d$ in the $z$-direction \emph{for all angles of incidence}, ideally,
\begin{equation}
\label{Eq:FreeSpace}
    \begin{aligned}
        &r_{-\frac{d}{2}\to -\frac{d}{2}}\left(\theta_0\right)\equiv0, &t_{-\frac{d}{2}\to \frac{d}{2}}\left(\theta_0\right)=e^{-jk_{z,0}d},
    \end{aligned}
\end{equation}
we enforce
\begin{equation}
\label{Eq:GrazingRequirement}
    \begin{aligned}
        &r_{-\frac{d}{2}\to -\frac{d}{2}}\left(\theta_0\to90^{\circ}\right)=0, &t_{-\frac{d}{2}\to \frac{d}{2}}\left(\theta_0\to90^{\circ}\right)=1
    \end{aligned}
\end{equation}
in Eq.\ (\ref{Eq:RT_Cascade}); note that this requirement is also compatible with the conclusion we have drawn in our discussion nearly towards the end of Sec.\ \ref{Sub:AllAngle}. Enforcing this analytical condition, we, in fact, find that there exists a unique closed-form solution, 
\begin{equation}
\label{Eq:TopBotRequirement}
    Y_{\mathrm{top}}=Y_{\mathrm{bot}}=-j\eta_0^{-1}\sqrt{\chi_{\mathrm{r}}}\tan\left(\frac{1}{2}k_0d\sqrt{\chi_{\mathrm{r}}}\right),
\end{equation}
to ensure null reflectance at grazing incidence, as requested by Eq.\ (\ref{Eq:GrazingRequirement}).

\begin{figure}
    \includegraphics[width=0.35\textwidth]{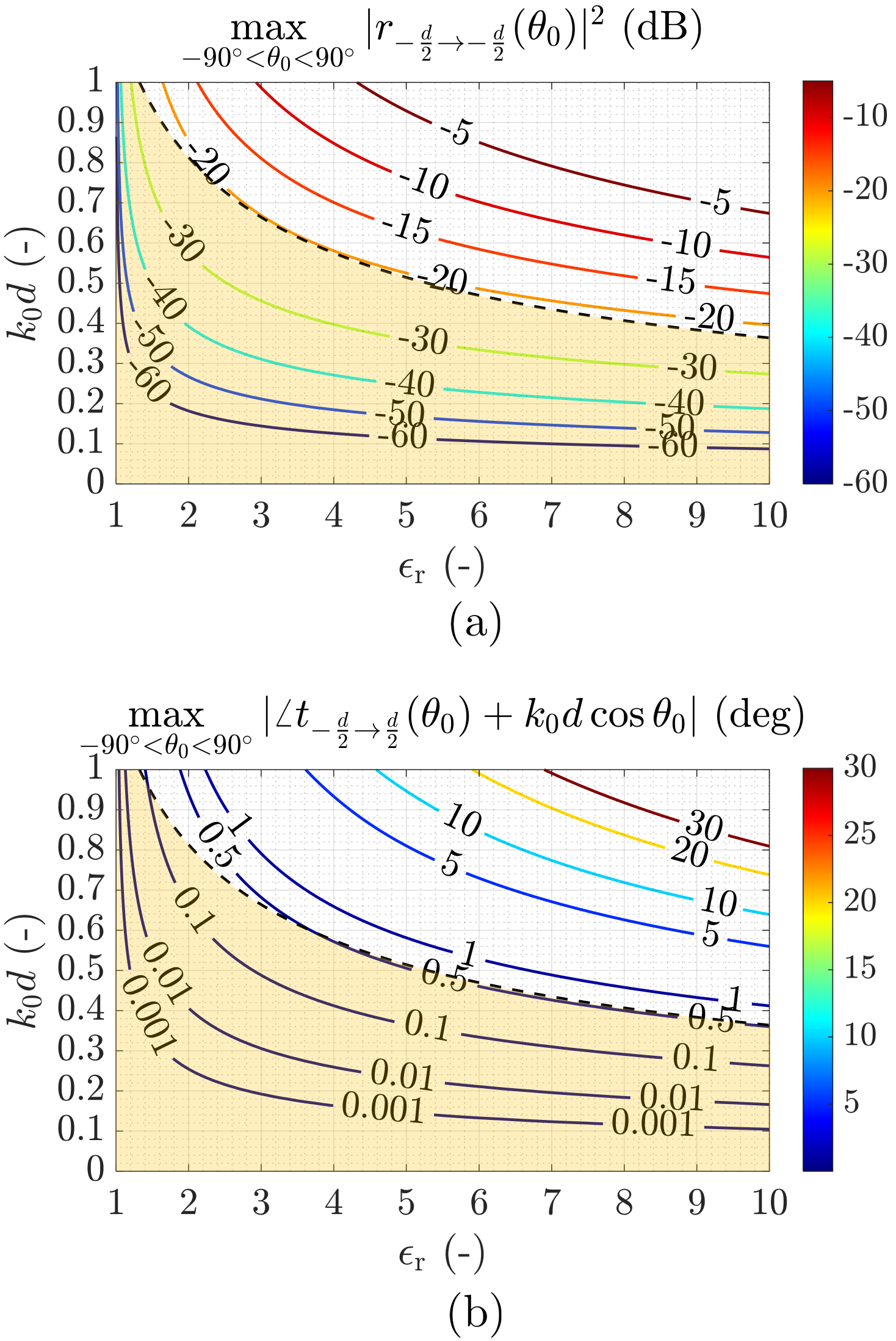}
\caption{(a) Maximal reflectance (dB) and (b) maximal absolute value of the phase error, relative to free space propagation, analytically calculated over the range $-90^{\circ}<\theta_0<90^{\circ}$ for a slab of relative permittivity $\epsilon_{\mathrm{r}}$ coated according to Eq.\ (\ref{Eq:TopBotRequirement}) (solid contour plots). Dashed black lines and light orange regions capture the region $k_0 d\sqrt{\epsilon_{r}}\leq1.15$, i.e., slab thicknesses of less than $0.183$ wavelengths in the dielectric ($\frac{\lambda_0}{\sqrt{\epsilon_{\mathrm{r}}}}$).}
\label{Fig:NumValid}
\end{figure}

Equipped by this straightforward solution for grazing-angle transparency, we analytically explore the entire angular response of the resultant configuration. We sweep the permittivity and normalized thickness values at the range $1<\epsilon_{\mathrm{r}}\leq 10$ and $0<k_{0}d\leq 1$; for each set of values $\epsilon_{\mathrm{r}}$ and $k_{0}d$ in question, we set the admittance values prescribed in Eq.\ (\ref{Eq:TopBotRequirement}) and plot the maximal power reflectance $\max\limits_{-90^{\circ}<\theta_0<90^{\circ}}|r_{-\frac{d}{2}\to -\frac{d}{2}}(\theta_0)|^2$ [dB units, Fig.\ \ref{Fig:NumValid}(a), solid contour plot] and maximal absolute value of the transmission-phase error relative to the goal phase behavior of free-space propagation [Eq.\ (\ref{Eq:FreeSpace})], $\max\limits_{-90^{\circ}<\theta_0<90^{\circ}}|\angle t_{-\frac{d}{2}\to \frac{d}{2}}(\theta_0)+k_0d\cos\theta_0|$ [Fig.\ \ref{Fig:NumValid}(b), solid contour plot], occurring at the range of $-90^{\circ}<\theta_0<90^{\circ}$, as predicted by Eq.\ (\ref{Eq:RT_Cascade}). We observe that the maximal power reflectance remains below $1\%$ ($<-20$ dB) and the phase error is less than $0.6^{\circ}$ for $k_{0}d\sqrt{\epsilon_{\mathrm{r}}}\lessapprox 1.15$ (formed by the dashed black line and light-orange filled area in Fig.\ \ref{Fig:NumValid}), i.e., when the overall substrate thickness $d$ is less than roughly 0.183 wavelengths in the dielectric ($\frac{\lambda_0}{\sqrt{\epsilon_{r}}}$). 

Remarkably, the above results reveal that the grazing-angle Huygens' condition manifested by Eq.\ (\ref{Eq:TopBotRequirement}) leads to a stronger universal outcome for thin slabs: not only does it overturn the unity reflectance for grazing incidence ($\theta_0\to90^{\circ}$), as intended, but it renders the slab practically transparent at \emph{all} angles of incidence as well ($-90^{\circ}<\theta_0<90^{\circ}$).

\begin{figure*}
    \includegraphics[width=\textwidth]{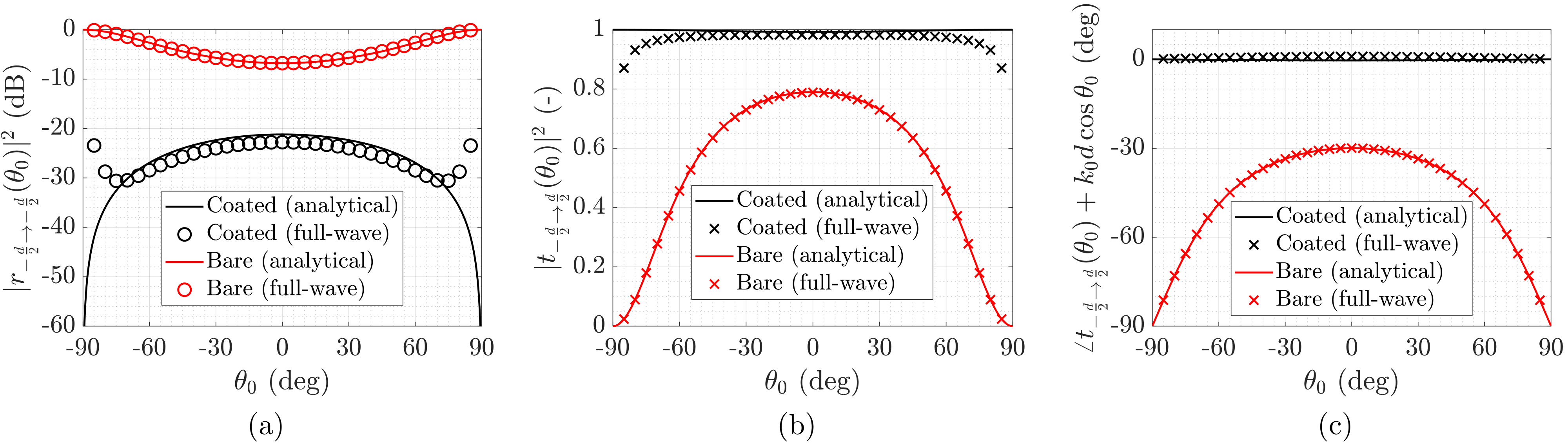}
\caption{(a)--(c)  Analytical [solid lines, Eq.\ (\ref{Eq:RT_Cascade})] and full-wave [recorded for the configuration in Fig.\ \ref{Fig:PCBConfig}(c); $\circ$ markers for reflectance and $\times$ markers for transmission coefficient] scattering parameters vs.\ angle of incidence $\theta_0$ for the coated (black) and bare (red) slabs of $d=60$-mil thick Rogers RO3003 at $f=20$ GHz: (a) reflectance (dB), (b) transmittance (linear units), and (c) transmission-phase error with respect to free-space propagation over distance $d$ along the $z$ direction.}
\label{Fig:PCBResults}
\end{figure*}

To demonstrate this interesting and powerful result and understand the underlying physical phenomena behind it, we design a practical case study  at $f=20$ GHz by coating a commercial Rogers RO3003 laminate of $\epsilon_{\mathrm{r}}=3$ and standard overall thickness $d=60$ mil$=1.524$ mm $\approx0.1016\lambda_0$ according to Eq.\ (\ref{Eq:TopBotRequirement}). For these parameters, the thickness compared to a wavelength in the dielectric, $\frac{d\sqrt{\epsilon_{r}}}{\lambda_0}\approx 0.176$, is well within the range of omnidirectional transparency once the slab is adequately coated (Fig. \ref{Fig:NumValid}). 

Equation (\ref{Eq:TopBotRequirement}) yields required admittance values of $Y_{\mathrm{top}}=Y_{\mathrm{bot}}=-0.6861j\eta_0^{-1}$. We set these values and inspect the reflectance $|r_{-\frac{d}{2}\to -\frac{d}{2}}\left(\theta_0\right)|^{2}$ [Fig. \ref{Fig:PCBResults}(a)], transmittance $|t_{-\frac{d}{2}\to \frac{d}{2}}(\theta_0)|^2$ [Fig.\ \ref{Fig:PCBResults}(b)], and transmission-phase error relative to free-space propagation $\angle t\left(\theta_0\right)+k_0d\cos\theta_0$ [Fig.\ \ref{Fig:PCBResults}(c)], as predicted by the analytical model of Eq.\ (\ref{Eq:RT_Cascade}) (solid lines), both for the coated slab (black) and for a reference scenario of a bare slab of identical permittivity and thickness (red, $Y_{\mathrm{top}}=Y_{\mathrm{bot}}=0$). The substantial reflectance off the bare slab, which achieves a minimal value of $\approx 21\%\approx -6.77$ dB at normal incidence and reaches unity (total reflection) at $\theta_0\to90^{\circ}$, is drastically reduced to below $0.76\%\approx -21.2$ dB across the entire angular range once the slab is coated; particularly, it rapidly drops to zero for $\theta_0\to 90^{\circ}$, as intended. The transmittance, which complements the reflectance to unity due to losslessnsess, closely follows similar dramatic improvement, remaining above $99.24\%$ for all angles. The transmission phase error for the bare slab, which approximately reads $-30^{\circ}$ at $\theta_0=0$ and worsens towards $-90^{\circ}$ as $\theta_0\to90^{\circ}$, is tremendously improved, upon coating the substrate, to $-0.36^{\circ}$ at $\theta_0=0$, while improving even further towards $0$ for $\theta_0\to90^{\circ}$, as intended.

Having obtained these excellent results and agreement with theory, which are simply achieved by merely employing the grazing-angle Huygens' condition, one may wonder whether such a MS-level design is related, in some sense, to an equivalent structure at the meta-atom level of Sec.\ \ref{Sec:MetaAtom}. In other words, can the PCB MS design in Fig.\ \ref{Fig:PCBConfig}(a) be described by means of effective meta-atom susceptibility values in the form of Eq.\ (\ref{Eq:SusceptComponents})? 

If so, one may expect that the effective electric susceptibility value $\widetilde{\chi}_{\mathrm{ee}}^{yy}$ obtained by such a query would coincide with the negative of the normal magnetic susceptibility $-\widetilde{\chi}_{\mathrm{mm}}^{zz}$, in concordance with the grazing-angle Huygens' condition [middle equality of Eq.\ (\ref{Eq:GenHuygensTE})], as preordained by our MS-level design procedure [Eqs.\ (\ref{Eq:GrazingRequirement}) and (\ref{Eq:TopBotRequirement})]. As for the tangential magnetic susceptibility $\widetilde{\chi}_{\mathrm{mm}}^{xx}$, we note that Huygens' condition for normal incidence [leftmost equality of Eq.\ (\ref{Eq:GenHuygensTE})], does not perfectly hold for the PCB MS in question (since its transmittance at this angle is slightly less than unity). Therefore, we expect a certain small discrepancy between the effective values of tangential electric $\widetilde{\chi}_{\mathrm{ee}}^{yy}$ and magnetic $\widetilde{\chi}_{\mathrm{mm}}^{xx}$ susceptibilities exhibited by the cascaded configuration.

To provide rigorous evidence that our hypothesis above actually holds and gather further essential insights, we characterize the coated MS configuration of Fig.\ \ref{Fig:PCBConfig}(a) according to the meta-atom-related method prescribed in \cite{Zaluski2016} (Appendix \ref{App:Char}). Indeed, the resultant electric susceptibility value, $\widetilde{\chi}_{\mathrm{ee}}^{yy}\approx0.0937$ excellently agrees with the negative of the normal magnetic susceptibility, $-\widetilde{\chi}_{\mathrm{mm}}^{zz}\approx0.0916$, in concurrence with the grazing-angle Huygens' condition at the meta-atom level.

Moreover, the effective tangential magnetic susceptibility reads $\chi_{\mathrm{mm}}^{xx}\approx-0.0809$, differing, to some extent, from the aforementioned value of $\widetilde{\chi}_{\mathrm{ee}}^{yy}$ and thus slightly deviating from Huygens' condition at normal incidence, as speculated. Nevertheless, such deviation is minuscule in the sense that the reflectance in this scenario still practically vanishes at all angles. This reduced sensitivity of Huygens' condition for normal incidence can be explained by investigating the reflectance in view of Eq.\ (\ref{Eq:scatTE}): provided that the grazing-angle Huygens' condition $\widetilde{\chi}_{\mathrm{ee}}^{yy}=-\widetilde{\chi}_{\mathrm{mm}}^{zz}$ is satisfied and that the susceptibility values are not exceedingly large, i.e., $|\widetilde{\chi}_{\mathrm{ee}}^{yy}|,|\widetilde{\chi}_{\mathrm{mm}}^{yy}|,|\widetilde{\chi}_{\mathrm{ee}}^{yy}+\widetilde{\chi}_{\mathrm{mm}}^{xx}|\ll 2$, the reflectance approximately reads $|r^{\mathrm{TE}}\left(\theta_0\right)|^{2}\approx 0.25\left(\widetilde{\chi}_{\mathrm{ee}}^{yy}-\widetilde{\chi}_{\mathrm{mm}}^{xx}\right)^{2}\cos^{2}\theta_0$; specifically, according to this approximation, the above susceptibility values yield maximal reflectance of $0.76\%\approx-21.2$ dB at normal incidence, in agreement with the black solid-line trace in Fig.\ \ref{Fig:PCBResults}(a). Hence, subject to small effective susceptibility values, as typically emulated by such thin cascades that adhere the grazing-angle Huygens' condition (evident by virtue of Fig.\ \ref{Fig:NumValid} and our exemplary discussion above), deviation from Huygens' condition for normal incidence negligibly affects the performance of all-angle transparency, as long as the grazing-angle Huygens' condition is satisfied.

These results further emphasize the dominance of the unconventional grazing-angle Huygens' condition in such reflectionless designs, compared to the insignificant sensitivity exhibited by the conventional Huygens' condition at normal incidence. Furthermore, they, in fact, reveal a powerful path to emulate and shape normal susceptibility components by means of simple multilayered stacks of tangentially polarizable inclusions. In other words, compared to standard HMS theory \cite{Pfeiffer2013Cascaded,Selvanayagam2013,Pfeiffer2013Millimeter,Wong2014,Glybovski2016,Chen2016,Epstein2016,Chen2018}, we find that not only do the coating admittance values control the effective tangential susceptibilities, but they simultaneously tune the normal ones as well. As discussed above, the heart of such equivalence is the combination of the various nonlocal features embedded in the structure. To conclude this overall analysis, these universal results show that thin dielectric slabs can be rendered fully transparent for \emph{all} angles of incidence by means of simple top and bottom admittance-sheet coatings designed via the grazing-angle Huygens' condition.

\subsection{Full-wave design and validation}
\label{Subsec:CascadeFullWave}
To demonstrate and validate our theoretical predictions in practice, we proceed with the case study considered above ($f=20$ GHz, $d=60$-mil thick RO3003 substrate), and realize the requested inductive admittance values ($Y_{\mathrm{top}}=Y_{\mathrm{bot}}=-0.6861j\eta_0^{-1}$) by printing meander-line copper strips on each facet \cite{Popov2019}, as depicted in Fig.\ \ref{Fig:PCBConfig}(c). We set a deep-subwavelength period, $L_{x}=3$ mm $\approx0.2\lambda_0$, along $x$ (abiding standard MS homogenization); a trace width of $w_{\mathrm{PCB}}=4$ mil $=0.1016$ mm, compatible with standard feature resolutions available in common PCB fabrication techniques; and a period of $L_{y}=4w_{\mathrm{PCB}}$ to allow continuous contact between meander-line segments accommodated by adjacent unit-cells. The meander width $W$ is reserved as a degree of freedom to tune the associated admittance values $Y_{\mathrm{top}}\left( W\right)=Y_{\mathrm{bot}}\left( W\right)$, for which we establish an accurate LUT, as follows.

\begin{figure}
    \includegraphics[width=0.35\textwidth]{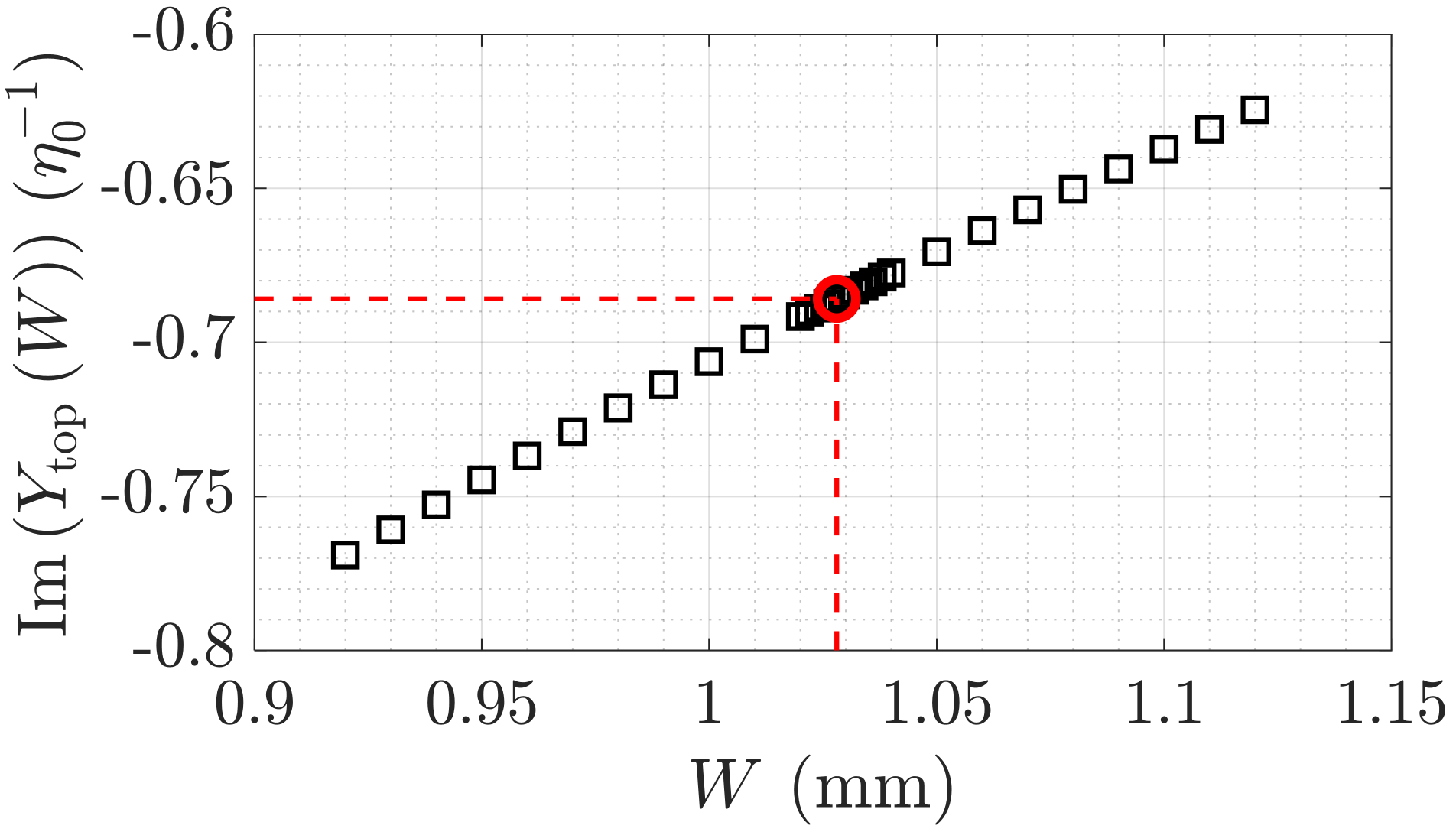}
\caption{Full-wave characterization at $\theta_0=85^{\circ}$ for the PCB configuration of Fig.\ \ref{Fig:PCBConfig}(c) with identical bottom and top meander-line width $W$ ($Y_{\mathrm{top}}\left(W\right)=Y_{\mathrm{bot}}\left( W\right)$, black square markers). The red $\circ$ marker shows the desired susceptance value to achieve the grazing-incidence Huygens' condition in our specific design, $\mathrm{Im}\left(Y_{\mathrm{top}}\right)=\mathrm{Im}\left(Y_{\mathrm{bot}}\right)\approx-0.6861\eta_0^{-1}$, which leads to $W=1.028$ mm.}
\label{Fig:PCBchar}
\end{figure}

For a given meander-width value $W$ in our range of interest, we model the composite of Fig.\ \ref{Fig:PCBConfig}(c) in CST and record the plane-wave reflection and transmission coefficients, $r_{-\frac{d}{2}\to -\frac{d}{2}}\left(\theta_0;W\right)$ and $t_{-\frac{d}{2}\to \frac{d}{2}}\left(\theta_0;W\right)$, under periodic boundary conditions, for $\theta_0=85^{\circ}$ (a near-grazing angle). Next, we follow Eq.\ (\ref{Eq:RT_Cascade}) and eliminate $Y_{\mathrm{top}}\left(W\right)$ in terms of the full-wave values of the scattering coefficients and the rest of the MS constituents,
\begin{equation}
\label{Eq:YtopChar}
    \begin{aligned}
        &Y_{\mathrm{top}}\left( W\right)=\frac{Y_{\mathrm{d}}}{j\sin\left(k_{z,\mathrm{d}}d\right)}\\
        &\times\left[\frac{1+r_{-\frac{d}{2}\to -\frac{d}{2}}\left(\theta_0;W\right)}{t_{-\frac{d}{2}\to \frac{d}{2}}\left(\theta_0;W\right)}-\cos\left(k_{z,\mathrm{d}}d\right)\right]-Y_{0}.
    \end{aligned}
\end{equation}

We repeat this process for several relevant values of $W$ and plot the obtained susceptance $\mathrm{Im}\left(Y_{\mathrm{top}}\left(W\right)\right)$ in Fig.\ \ref{Fig:PCBchar} (black square markers). Next, we utilize this LUT to tune $W=1.028$ mm and achieve the desired susceptance value [red $\circ$ marker and dashed lines in Fig.\ \ref{Fig:PCBchar}]. We thus satisfy the requirement to accomplish omnidirectional transparency and finalize the design procedure.

We probe the performance of this PCB prototype via full-wave simulations in CST. We illuminate the coated slab and a reference bare slab of the same material and dimensions with a $20$-GHz plane wave impinging at angle $-85^{\circ}\leq\theta_0\leq85^{\circ}$, and compare the full-wave scattering results to their analytical counterparts of Sec.\ \ref{Subsec:CascadeGHC}, as presented in Figs.\ \ref{Fig:PCBResults}(a)--(c). Indeed, the large full-wave reflectance [Fig.\ \ref{Fig:PCBResults}(a), $\circ$ markers] off the bare slab (red), which reaches a minimal value of $|r\left(\theta_0=0\right)|^{2}\approx21\%\approx -6.77$ dB and worsens towards unity at $\theta_0\to 90^{\circ}$, is drastically reduced to below $0.53\%\approx-22.77$ dB for all the angles in the inspected range once the slab is coated, exhibiting excellent agreement with theory. Likewise, the full-wave transmittance $|t_{-\frac{d}{2}\to\frac{d}{2}}\left(\theta_0\right)|^2$ [Fig.\ \ref{Fig:PCBResults}(b), $\times$ markers] follows similar congruence with theory, retaining values greater than $93.1\%$ for the coated slab at $-80^{\circ}\leq\theta_0\leq80^{\circ}$. Excellent correspondence between the full-wave and theoretical results of the phase error $\angle t_{-\frac{d}{2}\to\frac{d}{2}}\left(\theta_0\right) + k_{0}d\cos\theta_0$ [Fig.\ \ref{Fig:PCBResults}(c), $\times$ markers] is observed as well, achieving values lower than $0.99^{\circ}$, in the entire angular range. Despite some minor reduction in transmission efficiency, which, evidently, occurs solely due to inevitable copper and dielectric loss inflicted by the physical device, the full-wave results above indeed confirm that our goal of all-angle transparency, both in magnitude and phase, is practically met, as theoretically predicted.

Repeating the characterization process \cite{Zaluski2016} (Appendix \ref{App:Char}) for the full-wave investigation (Fig.\ \ref{Fig:PCBResults}) of the PCB prototype in Fig.\ \ref{Fig:PCBConfig}(c), yields the following effective susceptibility values: $\widetilde{\chi}_{\mathrm{ee}}^{yy}\approx0.0563$, $\widetilde{\chi}_{\mathrm{mm}}^{xx}\approx-0.09$, and $\widetilde{\chi}_{\mathrm{mm}}^{zz}\approx-0.0563$. These values are slightly shifted from those previously obtained for the analytical admittance-sheet cascades, mostly due to inevitable small loss present for this realistic design. Nonetheless, the persistent coincidence of the extracted electric susceptibility value ($\widetilde{\chi}_{\mathrm{ee}}^{yy}$) with that of the normal magnetic one ($\widetilde{\chi}_{\mathrm{mm}}^{zz}$) provides another essential evidence to the congruence of this design with the grazing-angle Hyugens' condition. Importantly, it further signifies the grazing angle as the natural pivot to the relation between the meta-atom and MS levels established by the nonlocal foundations discussed above. Subject to practical yet reasonable deviations, this excellent performance and agreement with theoretical predictions, with respect to all figures of merit, validates our framework above and allows us to proceed to more advanced inspections, as follows.

\begin{figure}
    \includegraphics[width=0.48\textwidth]{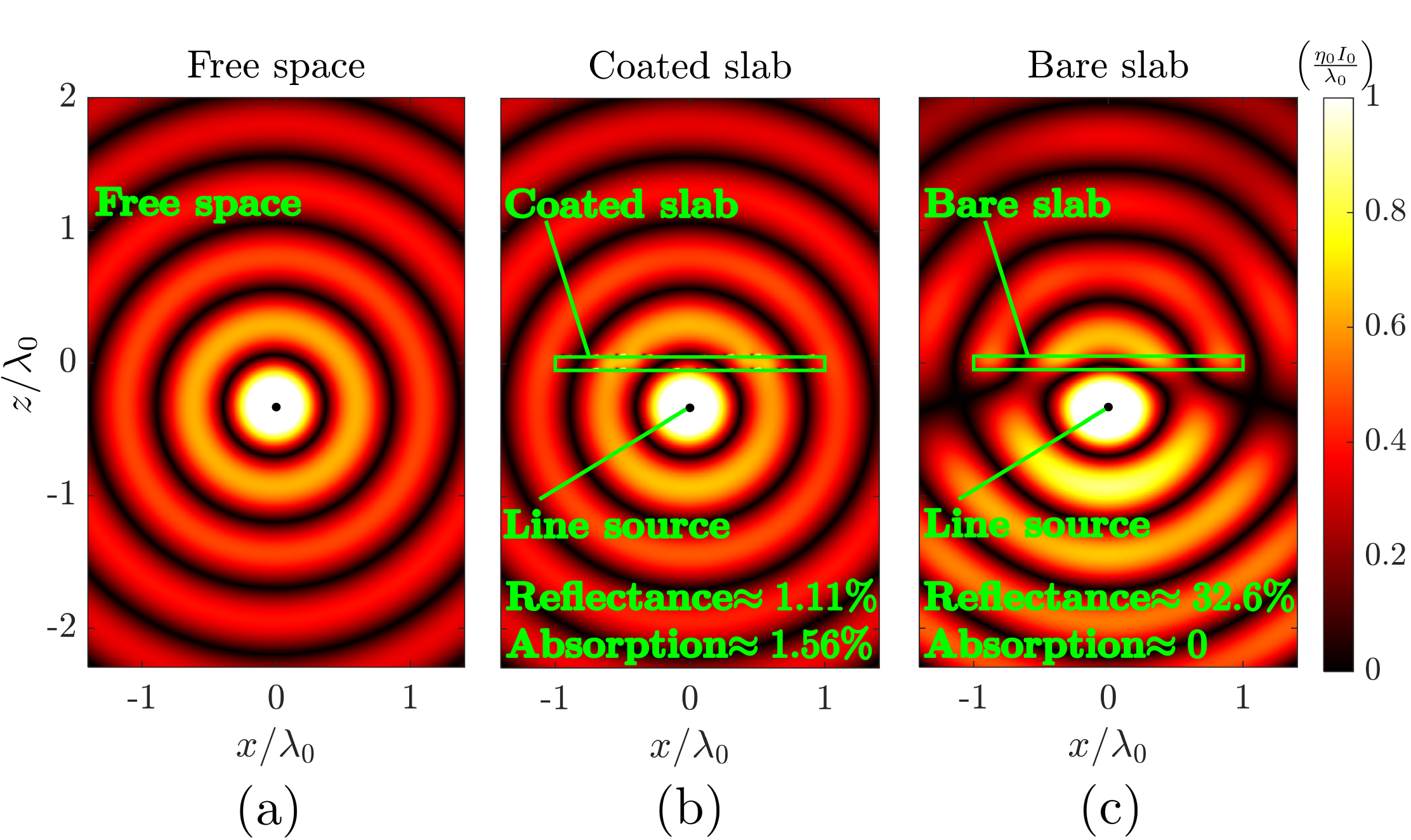}
\caption{2D distribution of $|\mathrm{Re}(E_{y}(x,z))|$ due to a line source of current $I_0$ at $(x,z)=(0,-\lambda_0/3)$: (a) free space (no slab); (b) with a coated slab of 2$\lambda_0$ (ten unit-cells) width along $x$ and infinite depth along $y$; (c) with a bare slab of the same dimensions and illumination.}
\label{Fig:PCBFields}
\end{figure}

Owing to the all-angle transparency of the MS, especially in the sense that its transmission phase practically coincides with that of propagation in free space for all angles of incidence, it is highly capable of preserving wavefronts. The reason is that each plane wave in the spectrum emanated off a practical source located near the MS is shifted across the MS with the same phase delay as it would have been had the MS been removed. To provide demonstration and further evidence for such notable functionality, we truncate the MS of Fig.\ \ref{Fig:PCBConfig}(c) along $x$ to ten periods (total width of $30$ mm $\approx2\lambda_0$), excite it with a 20-GHz $y$-directed line source of current $I_{0}$ located at $\left(x,z\right)=\left(0,-\lambda_0/3\right)$, and monitor the transverse field $|\mathrm{Re}\left(E_{y}\left(\vec{r}\right)\right)|$ on the $y=0$ plane in CST \footnote{The 2D TE configuration in this simulation (i.e., periodicity along $y$) is emulated by two boundary PEC planes located at $y=-2w_{\mathrm{PCB}}$ and $y=2w_{\mathrm{PCB}}$ (not shown), while the rest of the boundaries are perfectly matched to eliminate any reflection off them and thus simulate infinite 2D open space.} (Fig.\ \ref{Fig:PCBFields}). Figure \ref{Fig:PCBFields}(a) shows perfect cylindrical wavefronts radially propagating outwards, as the line source is allowed to radiate in free space without the slab; Fig.\ \ref{Fig:PCBFields}(b) shows the fields for the same setup, subject to the presence of the coated slab (green rectangle); and Fig.\ \ref{Fig:PCBFields}(c) depicts the fields when the coated slab is replaced by a bare slab (green rectangle) of the same material and dimensions.

Clearly, the bare slab inflicts substantial reflection, as evident from the interference pattern below the slab in Fig.\ \ref{Fig:PCBFields}(c). Moreover, it deforms the transmitted wavefronts above it [compared to free-space propagation in Fig.\ \ref{Fig:PCBFields}(a)] due to the undesired inherent angular behavior of the transmission phase $\angle t_{-\frac{d}{2}\to \frac{d}{2}}\left(\theta_0\right)$ in Fig.\ \ref{Fig:PCBResults}(c). However, as expected, once the slab is coated [Fig.\ \ref{Fig:PCBFields}(b)], the interference pattern and wavefornt distortion are drastically mitigated, such that the fields essentially coincide with those of the reference scenario of free space [Fig.\ \ref{Fig:PCBFields}(a)]. As a more quantitative measure for the MS performance under such practical excitation conditions, we estimate the total reflectance off the coated and bare slabs from CST (Appendix \ref{App:RefAbsorbCalc}). Indeed, the bare slab's poor performance of $\approx 32.6\%$ reflectance is dramatically improved to $\approx1.11\%$, with negligible absorption of $\approx 1.56\%$. These excellent results, both in terms of performance and agreement with theory, evidently conclude our careful procedure of full-wave design and validation, strongly verifying the omnidirectional transparency attributed to the coated slab.

\begin{figure}
    \includegraphics[width=0.45\textwidth]{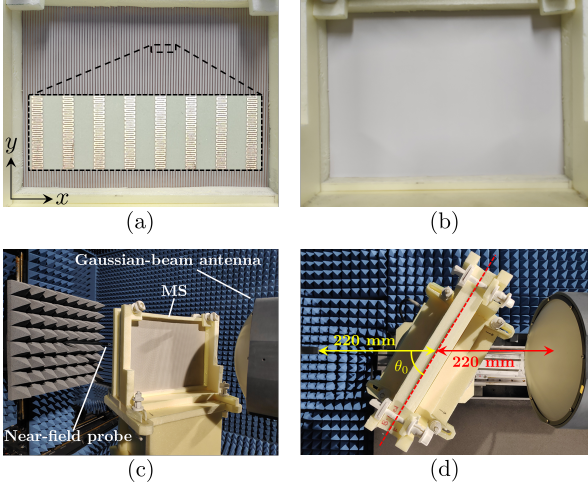}
\caption{Fabricated (a) coated (inset: closer view) and (b) bare slabs, as designed in Fig.\ \ref{Fig:PCBConfig}(c). (c) Perspective view of the experimental setup: the MS is approximately located at the focus of the Guassian-beam illumination from the antenna, while the near-field probe performs a planar scan. (d) Top view of the experimental setup. The angle of incidence $\theta_{0}$ is set with the help of a goniometer with $0.5^{\circ}$ resolution.}
\label{Fig:ExpSetup}
\end{figure}

\begin{figure}
    \includegraphics[width=0.41\textwidth]{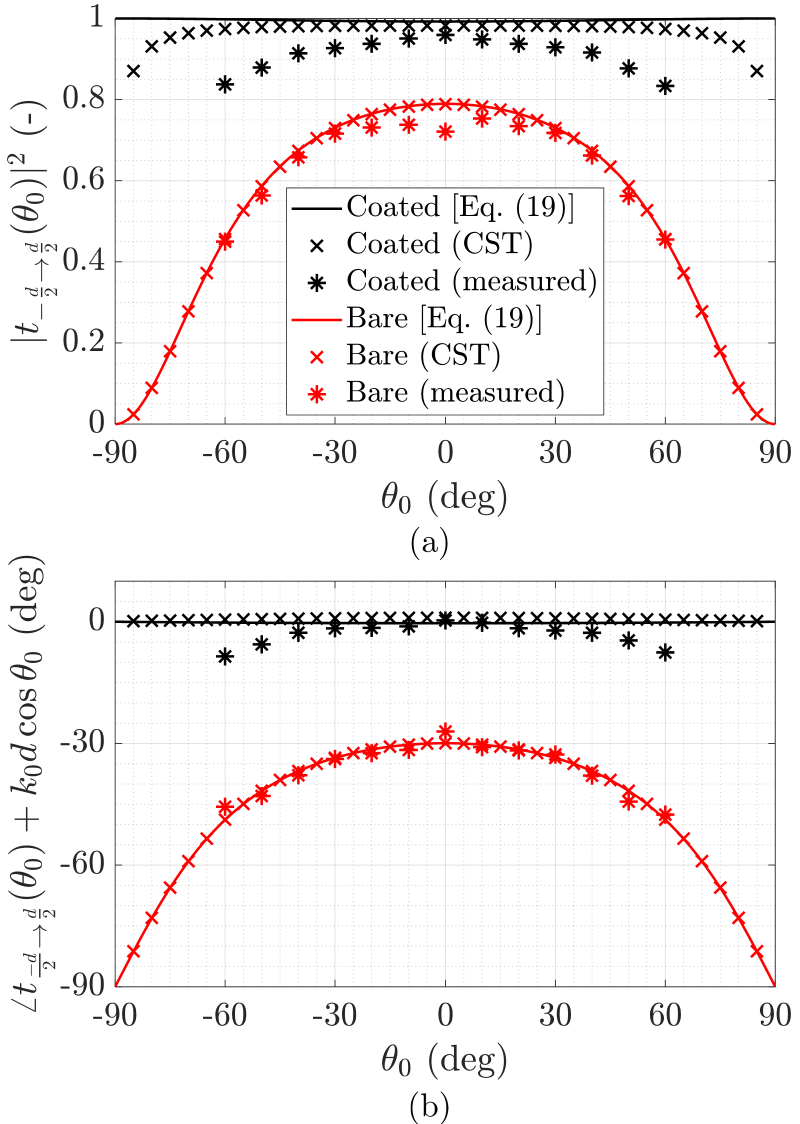}
\caption{Analytical (solid lines), full-wave ($\times$ markers), and measured ($*$ markers) for the (a) transmittance and (b) transmission-phase error vs.\ angle for the coated (black) and bare (red) slabs at $f=20$ GHz.}
\label{Fig:ExpRes20}
\end{figure}

\subsection{Experimental measurements}
In order to experimentally verify our theoretical observations, the PCB-compatible design from Fig.\ \ref{Fig:PCBConfig}(c) was fabricated on a commercial 9''$\times$12'' ($\approx15.24\lambda_0$ along the meander $y$-direction and $\approx20.32\lambda_0$ along the $x$-direction) Rogers RO3003\textsuperscript{TM} laminate [Fig.\ \ref{Fig:ExpSetup}(a)] and characterized via through measurements in the anechoic chamber at the Technion [Figs.\ \ref{Fig:ExpSetup}(c) and (d)].
The prototype was mounted on a rotatable foam holder, placed approximately at the focus of a Gaussian-beam antenna (Millitech Inc., GOA-42-S000094, focal distance of 196 mm $\approx 13\,\lambda$), illuminating the device under test (DUT) from a distance of $220$ mm with a quasi-planar wavefront.
A planar near-field measurement system (MVG/Orbit-FR) was utilized to record the forward scattering pattern, by scanning an area of $400\times400$ mm\textsuperscript{2} ($\approx 26.6\lambda_0\times26.6\lambda_0$) in the $x$ and $y$ directions, at a distance of $220$ mm from the center of the MS plane, from which the far-field pattern can be deduced using the equivalence principle \cite{Balanis2012}.

After setting the angle of incidence $\theta_{0}$ by rotation of the foam holder, an 18--22-GHz sweep was performed, accompanied by a reference measurement, for which the MS was removed and the Gaussian-beam antenna directly illuminated the same near-field scanning plane. By repeating this procedure for a desired set of angles covering the range of $-60^{\circ}\leq\theta_{0}\leq60^{\circ}$ (limited by practical setup considerations) the angular response of the MS transmission can be obtained. The magnitude and phase of the forward far-field gain, which are calculated from the near-field measurements and calibrated with respect to the reference measurements, represent the power transmittance $|t_{-\frac{d}{2}\to \frac{d}{2}}\left(\theta_0\right)|^{2}$ and the transmission-phase error with respect to free-space propagation $\angle t_{-\frac{d}{2}\to \frac{d}{2}}\left(\theta_0 \right)+k_{0}d\cos\theta_0$. For comparison, identical procedure is repeated for another specimen of a bare Rogers RO3003 slab of the same dimensions [Fig.\ \ref{Fig:ExpSetup}(b)].

Figure \ref{Fig:ExpRes20}(a) compares the analytical (solid lines), full-wave ($\times$ markers), and measured ($*$ markers) transmittance vs.\ angle of incidence at $f=20$ GHz for both the coated (black) and bare (red) slabs. The same comparison is performed for the transmission-phase error in Fig.\ \ref{Fig:ExpRes20}(b). Overall, the measured results agree reasonably well with their full-wave predictions, while exhibiting increasing deviations as the angle of incidence is increased. Despite such discrepancies, we observe very good improvement of measured transmittance for the entire angular range, e.g., from $\approx72.1\%$ to $\approx95.9\%$ (improvement by factor of $\approx 1.26$) at normal incidence and from $\approx45\%$ to $\approx83.5\%$ (factor of $\approx1.86$) at $\theta_0=\pm 60^{\circ}$, and of phase error, e.g, from $\approx-27^{\circ}$ to $\approx0.33^{\circ}$ at normal incidence and from $\approx -46^{\circ}$ to $\approx -8^{\circ}$ for $\theta_0= \pm 60^{\circ}$. Hence, this experiment confirms that the coated slab designed according to the proposed GHC-inspired scheme in Sec. \ref{Subsec:CascadeGHC} functions well, i.e., dramatically mitigates reflectivity and phase aberrations compared to its bare counterpart over a wide range of angles.

Indeed, the measured performance of the coated slab is, to some minor extent, less efficient than theoretically expected. This behavior can be caused by possible fabrication inaccuracies, which may slightly alter the substrate properties or the coating admittance values and, hence, lead to certain deviation from the GHC-inspired condition [Eq.\ (\ref{Eq:TopBotRequirement})], thus deterring ideal all-angle transparency. In particular,
the typical tolerance values indicated in \cite{Rautio2011,Rabinovich2019,Budiana2020} for the dielectric constant $\epsilon_{\mathrm{r}}$ and slab thickness $d$ can reasonably explain the discrepancies and trends observed in our measured results.

\begin{figure}
    \includegraphics[width=0.48\textwidth]{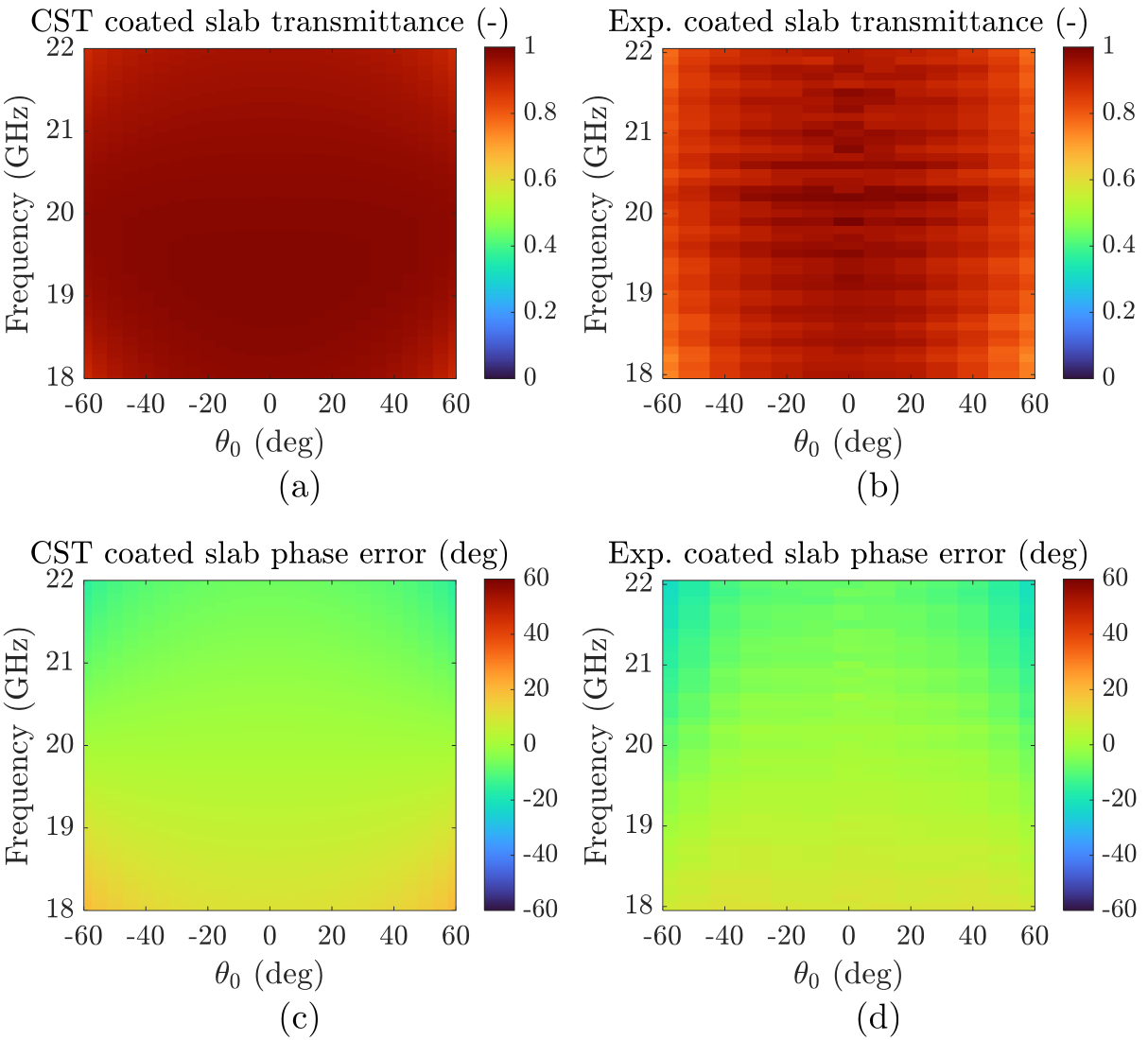}
\caption{Full-wave [(a) and (c)] and experimental [(b) and (d)] results for the transmittance [(a) and (b)] and transmission-phase error [(c) and (d)] vs. angle and frequency for the coated slab.}
\label{Fig:ExpResFreq}
\end{figure}

Additional factors, such as tolerances in other design parameters and the finite spatial spectral width of the Gaussian beam, which includes components associated with near-grazing angles of large reflectance, may further complement this study in a way that describes the measured phenomena more accurately. Overall, based on this investigation, we deduce that the good performance presented by the fabricated device matches well to the theoretical expectations, subject to reasonable practical manufacturing tolerances. This further validates the viability of our design concept relying on the grazing-angle Huygens' condition and the GHC in general.

Before concluding, we compare in Fig.\ \ref{Fig:ExpResFreq} between the full-wave [(a) and (c)] and measured [(b) and (d)] frequency response of transmittance [(a) and (b)] and transmission-phase error [(c) and (d)] for the coated MS. The experimental results match well to the full-wave expectations. Despite the absence of apriori bandwidth considerations in our GHC-inspired design procedure, both full-wave and measured data present relatively broadband operation (Appendix \ref{App:Freq}). Specifically, simulated results show that the composite retains near-unity transmittance, $|t_{-\frac{d}{2}\to\frac{d}{2}}\left(\theta_0 \right)|^{2}>-0.5$ dB ($\approx89.1\%$), and relatively low phase error, $|\angle t_{-\frac{d}{2}\to \frac{d}{2}}\left(\theta_0 \right)+k_{z,0}d|<20^{\circ}$ for $-60^{\circ}\leq\theta_0\leq60^{\circ}$ at 18--22 GHz (20\% fractional bandwidth). Summarizing our experimental findings above, we highlight that our measurements indeed verify the simulated results of omnidirectional transparency inspired by the GHC within a respectable bandwidth, subject to reasonable practical fabrication inaccuracies.

\section{Conclusion}
\label{sec:Conclusion}
To conclude, we have formulated and validated a profound generalization of Huygens' condition, that is, the GHC, to facilitate planar all-angle reflectionless MS devices, both at the meta-atom and MS levels of design. Deducing that such condition is simply achieved by setting null backscattering at normal and grazing incidence, we have revealed that the latter establishes the key insight to resolve the long standing issue of singular wave-impedance values related to Fresnel reflection. In this regard, we have unraveled the role of each susceptibility component, particularly the normal one, in the overall MS angular response and its fundamental relation to the nonlocal mechanisms that govern such spatial dispersion.

While a few previous reports, e.g., \cite{Radi2015,Im2018}, have formulated mathematical conditions for all-angle transparency in terms of angularly dependent constituents,
a key universal principle easily and intuitively transferable to realistic MS theory and implementation of simple local responses has not been provided.
Herein, the fundamental GHC paradigm fills this gap and elucidates the profound physical aspects indispensable to such a functionality, in particular, the intricate role of grazing incidence and its elemental meta-atom relation to local normal response.

Harnessing these insights in a practical MS-level design, we have rigorously rendered a thin dielectric slab omnidirectionally transparent by means of simple admittance coatings compatible with standard PCB technology. In particular, we have achieved this functionality based on the newfound grazing-angle Huygens' condition \emph{per se}, which leads to a simple closed-form analytical solution, rather than extensive and time-consuming full-wave optimizations. Importantly, we have demonstrated how the evidently universal framework of the effective susceptibilities consistently captures the essence of both the meta-atom and MS realizations, which rely on intricate nonlocal phenomena of diverse origins. In this specific aspect, we have, in effect, provided a valuable generalization to recent homogenization concepts studied for elementary configurations of thin sheets \cite{Tiukuvaara2022}, clarified their underlying nonlocal foundations, and shown how they can be controlled by simple practical means. Apropos to the effective susceptibility extraction and analysis shown herein, we believe that the GHC may be considered in the future as a relevant performance criterion or touchstone of all-angle transparency to be embedded in other design schemes and quantitatively assess their performance; for instance, it can be readily adopted to explore other previous devices, relying on different engineering considerations, such as \cite{Radi2015,Im2018,Goshen2023}.

Subject to reasonable practical fabrication and measurement inaccuracies, our design exhibits very good performance, confirming that this solution can be readily utilized by itself, e.g., as an efficient low-profile planar radome for wide-angle beam-steering applications. Intriguingly, the unconventional, yet simply applied, nonlocal paradigm presented in this paper reveals that multilayered PCBs of elementary tangential response can be harnessed to emulate normal susceptibilities; hence, it indeed proves promising capabilities as a fundamental convention to be incorporated in future designs of advanced wave-manipulating components, such as flat lenses, space-squeezing plates, and analog optical computing devices for image processing.

\section*{Acknowledgments}
This work was supported by the 2022 IEEE Antennas and Propagation Society Fellowship (APSF). The authors gratefully thank Rogers Corporation for providing one of the PCB samples measured in this study. We also thank Y.\ Milyutin and A.\ Cohen of the Micro and Nano Fabrication Unit (MNFU) at the Technion for their kind technical assistance in preparing this laminate for measurement. We would also like to acknowledge D.\ Dikarov of the Communication Laboratory of the Electrical and Computer Engineering Faculty at the Technion for his kind administrative assistance related to fabricating the specimens measured in this paper.

\appendix

\section{Scattering coefficient expressions for the TM-polarized scenario}
\label{App:TMexpressions}
For the TM-polarized scenario in Sec.\ \ref{Sec:GHC}, the expressions for the scattering coefficients defined in Eq.\ (\ref{Eq:Hy}) are found to have the same form of Eqs.\ (\ref{Eq:scatTE}) and (\ref{Eq:TECoeff}), subject to the substitutions indicated in in Eq.\ (\ref{Eq:Substitutions}), i.e.,
\begin{equation}
\label{Eq:scatTM}
    \begin{aligned}
    r^{\mathrm{TM}}\left(\theta_0\right)&=\frac{r^{\mathrm{TM}}_{0}+r^{\mathrm{TM}}_{2}\widetilde{k}_{z,0}^{2}}{d^{\mathrm{TM}}_{0}+d^{\mathrm{TM}}_{1}\widetilde{k}_{z,0}+d^{\mathrm{TM}}_{2}\widetilde{k}_{z,0}^{2}+d^{\mathrm{TM}}_{3}\widetilde{k}_{z,0}^{3}},\\
    t^{\mathrm{TM}}\left(\theta_0\right)&=\frac{t^{\mathrm{TM}}_{1}\widetilde{k}_{z,0}+t^{\mathrm{TM}}_{3}\widetilde{k}_{z,0}^{3}}{d^{\mathrm{TM}}_{0}+d^{\mathrm{TM}}_{1}\widetilde{k}_{z,0}+d^{\mathrm{TM}}_{2}\widetilde{k}_{z,0}^{2}+d^{\mathrm{TM}}_{3}\widetilde{k}_{z,0}^{3}},
    \end{aligned}
\end{equation}
where
\begin{equation}
\label{Eq:TMCoeff}
    \begin{aligned}
        &\begin{aligned}
            &r^{\mathrm{TM}}_0=-2\left( \widetilde{\chi}_{\mathrm{mm}}^{yy}+\widetilde{\chi}_{\mathrm{ee}}^{zz}\right),
            &r^{\mathrm{TM}}_2=2\left( \widetilde{\chi}_{\mathrm{ee}}^{xx}+\widetilde{\chi}_{\mathrm{ee}}^{zz}\right),
        \end{aligned}\\
        &t^{\mathrm{TM}}_1=-j\left[\right(\widetilde{\chi}_{\mathrm{mm}}^{yy}+\widetilde{\chi}_{\mathrm{ee}}^{zz}\left) \widetilde{\chi}_{\mathrm{ee}}^{xx}+4\right],\\
        &\begin{aligned}
            &t^{\mathrm{TM}}_3=j \widetilde{\chi}_{\mathrm{ee}}^{xx}\widetilde{\chi}_{\mathrm{ee}}^{zz},
            &d^{\mathrm{TM}}_0=2\left( \widetilde{\chi}_{\mathrm{mm}}^{yy}+\widetilde{\chi}_{\mathrm{ee}}^{zz}\right),
        \end{aligned}\\
        &d^{\mathrm{TM}}_{1}=j\left[ \left( \widetilde{\chi}_{\mathrm{mm}}^{yy}+\widetilde{\chi}_{\mathrm{ee}}^{zz}\right)\widetilde{\chi}_{\mathrm{ee}}^{xx}-4\right],\\
        &\begin{aligned}
            &d^{\mathrm{TM}}_2=2\left( \widetilde{\chi}_{\mathrm{ee}}^{xx}-\widetilde{\chi}_{\mathrm{ee}}^{zz}\right),
            &d^{\mathrm{TM}}_3&=-j\widetilde{\chi}_{\mathrm{ee}}^{xx}\widetilde{\chi}_{\mathrm{ee}}^{zz}.
        \end{aligned}
    \end{aligned}
\end{equation}

\section{Meta-atom characterization}
\label{App:Char}

\begin{figure*}
    \includegraphics[width=0.83\textwidth]{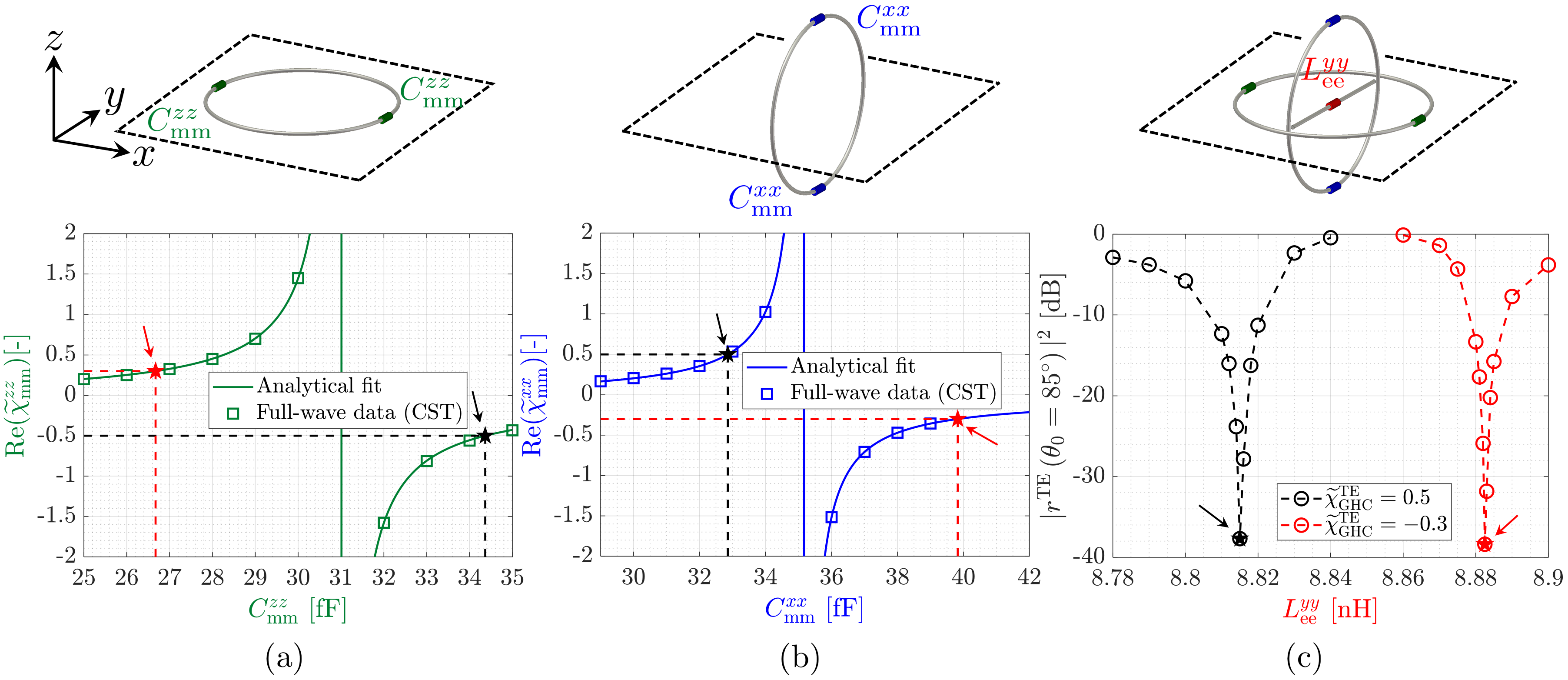}
\caption{Systematic design of the meta-atom from Fig.\ \ref{Fig:MetaAtom} at $f=20$ GHz: (a) an individual in-plane capacitively loaded loop ($\widetilde{\chi}_{\mathrm{mm}}^{zz}$); (b) an individual perpendicular capacitively loaded loop ($\widetilde{\chi}_{\mathrm{mm}}^{xx}$); (c) finalized configuration, in which both the loops in (a) and (b) are placed together with a straight inductively loaded wire ($\widetilde{\chi}_{\mathrm{ee}}^{yy}$). Look-up tables relating the susceptibility and capacitance values for the inclusions in (a) and (b), appear below their schematics, respectively: square markers represent full-wave characterization results based on \cite{Zaluski2016}, which are fitted by an approximated analytical formula based on \cite{Tretyakov2003,Shaham2021} (solid lines); pentagram ($\star$) markers stand for the chosen values for $\widetilde{\chi}_{\mathrm{GHC}}^{\mathrm{TE}}=0.5$ (black) and $\widetilde{\chi}_{\mathrm{GHC}}^{\mathrm{TE}}=-0.3$ (red) [Table \ref{Tab:FinalVals}]. (c) Finalization of the design for $\widetilde{\chi}_{\mathrm{GHC}}^{\mathrm{TE}}=0.5$ (black) and $\widetilde{\chi}_{\mathrm{GHC}}^{\mathrm{TE}}=-0.3$ (red): for each of the chosen capacitance values in (d) and (e), the inductance $L_{\mathrm{ee}}^{yy}$ in (c) is swept while the reflectance $|r^{\mathrm{TE}}\left(\theta_0=85^{\circ}\right)|^{2}$ at $\theta_0=85^{\circ}$ is recorded in CST ($\circ$ markers and dashed lines); pentagram markers show the selected values [Table \ref{Tab:FinalVals}] of minimum (practically vanishing) reflectance values.}
\label{Fig:MetaAtomChar}
\end{figure*}

\begin{table}
\centering
\begin{tabular}{l | c | c}
	& $\widetilde{\chi}_{\mathrm{GHC}}^{\mathrm{TE}}=0.5$ & $\widetilde{\chi}_{\mathrm{GHC}}^{\mathrm{TE}}=-0.3$\\
	\hline\hline
    $\widetilde{\chi}_{\mathrm{mm}}^{zz}$ [-] &-0.5 &0.3\\
	$C_{\mathrm{mm}}^{zz}$  [fF] &34.365 &26.674\\
    \hline
    $\widetilde{\chi}_{\mathrm{mm}}^{xx}$ [-] &0.5 &-0.3\\
    $C_{\mathrm{mm}}^{xx}$  [fF] &32.858 &39.826\\
    \hline
    $\widetilde{\chi}_{\mathrm{ee}}^{yy}$ [-] &0.5 &-0.3\\
	$L_{\mathrm{ee}}^{yy}$  [nH] &8.815 &8.8825\\
    \hline
\end{tabular}
\caption{Chosen final values for each of the designs, $\widetilde{\chi}_{\mathrm{GHC}}^{\mathrm{TE}}=0.5$ and $\widetilde{\chi}_{\mathrm{GHC}}^{\mathrm{TE}}=-0.3$, as selected by the pentagram markers ($\star$) in Figs.\ \ref{Fig:MetaAtomChar}(a)--(c).}
\label{Tab:FinalVals}
\end{table}

Given the scattering properties of a MS unit-cell (realistic or abstract), one may equivalently describe it via effective susceptibility values, which reproduce its functionality when substituted in the GSTCs [Eqs.\ (\ref{Eq:GSTCs}) and (\ref{Eq:Suscept})]. Assuming that the susceptibility components corresponding to the MS under inspection comply with Eq.\ (\ref{Eq:SusceptComponents}), Ref.\ \cite{Zaluski2016} has suggested the following characterization scheme to extract such effective values.

For the TE-polarized scenario, the field reflection and transmission coefficients of the unit cell are first recorded (under periodic boundary conditions) at normal incidence, $r^{\mathrm{TE}}(0)$ and $t^{\mathrm{TE}}(0)$, to yield the effective tangential electric and magnetic susceptibilities via
\begin{equation}
\label{Eq:CharTangential}
    \begin{aligned}
        \widetilde{\chi}_{\mathrm{ee}}^{yy}&=2j\frac{r^{\mathrm{TE}}\left(0\right)+t^{\mathrm{TE}}\left(0\right)-1}{r^{\mathrm{TE}}\left(0\right)+t^{\mathrm{TE}}\left(0\right)+1},\\
        \widetilde{\chi}_{\mathrm{mm}}^{xx}&=2j\frac{t^{\mathrm{TE}}(0)-r^{\mathrm{TE}}\left(0\right)-1}{t^{\mathrm{TE}}\left(0\right)-r^{\mathrm{TE}}\left(0\right)+1}.
    \end{aligned}
\end{equation}
Characterization of the normal magnetic component, in addition to the tangential ones in Eq.\ (\ref{Eq:CharTangential}), necessitates additional reflection and transmission data, $r^{\mathrm{TE}}\left(\theta_{\mathrm{c,mm}}^{zz}\right)$ and $t^{\mathrm{TE}}\left(\theta_{\mathrm{c,mm}}^{zz}\right)$, for yet another incidence scenario at an oblique angle $\theta_{\mathrm{c,mm}}^{zz}\neq 0$, such that
\begin{equation}
\label{Eq:CharNormal}
    \begin{aligned}
    &\widetilde{\chi}_{\mathrm{mm}}^{zz}\left(\theta_{\mathrm{c,mm}}^{zz}\right)=-\frac{\widetilde{\chi}_{\mathrm{ee}}^{yy}}{\sin^{2}\theta_{\mathrm{c,mm}}^{zz}}\\
    &-2j\frac{\cos\theta_{\mathrm{c,mm}}^{zz}}{\sin^{2}\theta_{\mathrm{c,mm}}^{zz}}\frac{1-r^{\mathrm{TE}}\left(\theta_{\mathrm{c,mm}}^{zz}\right)-t^{\mathrm{TE}}\left(\theta_{\mathrm{c,mm}}^{zz}\right)}{1+r^{\mathrm{TE}}\left(\theta_{\mathrm{c,mm}}^{zz}\right)+t^{\mathrm{TE}}\left(\theta_{\mathrm{c,mm}}^{zz}\right)}.
    \end{aligned}
\end{equation}
A similar scheme for TM-polarized fields and susceptibility components ($\widetilde{\chi}_{\mathrm{ee}}^{xx}$, $\widetilde{\chi}_{\mathrm{mm}}^{yy}$, and $\widetilde{\chi}_{\mathrm{ee}}^{zz}$) can be separately deduced by following the dual replacements in Eq.\ (\ref{Eq:Substitutions}) and Appendix \ref{App:TMexpressions}.

To realize a generalized Huygens' MS with a desired common susceptibility value $\widetilde{\chi}_{\mathrm{GHC}}^{\mathrm{TE}}$ by means of the meta-atom in Fig.\ \ref{Fig:MetaAtom}, our goal is to tune the load values ($C_{\mathrm{mm}}^{zz}$, $C_{\mathrm{mm}}^{xx}$, and $L_{\mathrm{ee}}^{yy}$) to accomplish effective susceptibility values ($\widetilde{\chi}_{\mathrm{mm}}^{zz}$, $\widetilde{\chi}_{\mathrm{mm}}^{xx}$, $\widetilde{\chi}_{\mathrm{ee}}^{yy}$) that meet with the GHC of Eq.\ (\ref{Eq:GenHuygensTE}). To this end, we first establish a LUT individually for the in-plane loop associating the capacitance value $C_{\mathrm{mm}}^{zz}$ with a corresponding value $\widetilde{\chi}_{\mathrm{mm}}^{zz}$ via \cite{Zaluski2016}. 

For a given capacitance value $C_{\mathrm{mm}}^{zz}$ under inspection, we model the loaded loop inclusion in Fig.\ \ref{Fig:MetaAtom} in CST and illuminate it from below ($z<0$) with a 20-GHz plane wave under periodic boundary conditions. The field reflection and transmission coefficients, are recorded both for normal, $r^{\mathrm{TE}}\left(0\right)$ and $t^{\mathrm{TE}}\left(0\right)$, and oblique, $r^{\mathrm{TE}}\left(\theta_{\mathrm{c,mm}}^{zz}\right)$ and $t^{\mathrm{TE}}\left(\theta_{\mathrm{c,mm}}^{zz}\right)$, incidence with $\theta_{\mathrm{c,mm}}^{zz}=30^{\circ}$. By substituting these values in Eqs.\ (\ref{Eq:CharTangential}) and Eqs.\ (\ref{Eq:CharNormal}), we find the normal magnetic susceptibility value of the loaded loop $\widetilde{\chi}_{\mathrm{mm}}^{zz}$ for the given capacitance $C_{\mathrm{mm}}^{zz}$. We repeat the process for several relevant values of $C_{\mathrm{mm}}^{zz}$ and plot the obtained $\mathrm{Re}\left(\widetilde{\chi}_{\mathrm{mm}}^{zz}\right)$ as square markers in Fig.\ \ref{Fig:MetaAtomChar}(a).

Next, to reduce the amount of full-wave simulations required to retrieve such LUT values, we fit the full-wave data with an approximate formula obtained by simple quasistatic lumped-circuit analysis \cite{Tretyakov2003,Shaham2021}, which associates the physical parameters of the (lossless) loop and its load with a susceptibility value,
\begin{equation}
\label{Eq:mmzzLoopSuscept}
  \mathrm{Re}\left(\widetilde{\chi}_{\mathrm{mm}}^{zz}\right)\approx\frac{\alpha_{\mathrm{mm}}^{zz} C_{\mathrm{mm}}^{zz}}{1-\beta_{\mathrm{mm}}^{zz} C_{\mathrm{mm}}^{zz}},
\end{equation}
where $\alpha_{\mathrm{mm}}^{zz}$ and $\beta_{\mathrm{mm}}^{zz}$ are frequency-dependent constants determined by the geometric and electromagnetic properties of the loop along with the unit-cell dimensions \cite{Shaham2021}. Following the method of least squares, this fit is shown as the solid line in Fig.\ \ref{Fig:MetaAtomChar}(a); the estimated coefficient values are $\alpha_{\mathrm{mm}}^{zz}\!\approx\!1.57\!\times\!10^{-3}$ $\mathrm{fF}^{-1}$ and $\beta_{\mathrm{mm}}^{zz}\!\approx\! 32.25\!\times\!10^{-3}$ $\mathrm{fF}^{-1}$. The tangentially polarizable loop in Fig.\ \ref{Fig:MetaAtom}, can be individually characterized in a likewise manner [Fig.\ \ref{Fig:MetaAtomChar}(b)], where the relation between the susceptibility $\widetilde{\chi}_{\mathrm{mm}}^{xx}$ and the capacitance $C_{\mathrm{mm}}^{xx}$ reads
\begin{equation}
\label{Eq:mmxxLoopSuscept}
    \mathrm{Re}\left(\widetilde{\chi}_{\mathrm{mm}}^{xx}\right)\approx\frac{\alpha_{\mathrm{mm}}^{xx} C_{\mathrm{mm}}^{xx}}{1-\beta_{\mathrm{mm}}^{xx} C_{\mathrm{mm}}^{xx}},
\end{equation}
where the estimated factors are $\alpha_{\mathrm{mm}}^{xx}\approx9.9835\times10^{-4}$ $\mathrm{fF}^{-1}$ and $\beta_{\mathrm{mm}}^{xx}\approx28.436\times10^{-3}$ $\mathrm{fF}^{-1}$).

To realize the typical values of $\widetilde{\chi}_{\mathrm{GHC}}^{\mathrm{TE}}=0.5$ (black markers and traces in Fig.\ \ref{Fig:MetaAtomChar}) and $\widetilde{\chi}_{\mathrm{GHC}}^{\mathrm{TE}}=-0.3$ (red markers and traces in Fig.\ \ref{Fig:MetaAtomChar}), which were taken for the demonstration in Sec.\ \ref{Sec:MetaAtom}, we fetch the capacitance values for each design from the LUTs constructed in Figs.\ \ref{Fig:MetaAtomChar}(a) and (b) [pentagram ($\star$) markers, see Table \ref{Tab:FinalVals} as well], set them in the compound meta atom in Fig.\ \ref{Fig:MetaAtom}, and sweep the inductance $L_{\mathrm{ee}}^{yy}$ while monitoring the reflectance at $\theta_0=85^{\circ}$ (a near-grazing angle) in CST, as shown in Fig.\ \ref{Fig:MetaAtomChar}(c). Lastly, we fix each inductance $L_{\mathrm{ee}}^{yy}$ to the value that minimizes such reflection (pentagram markers in Fig.\ \ref{Fig:MetaAtomChar}(c) and values in Table \ref{Tab:FinalVals}). This concludes our design procedure in Sec.\ \ref{Sec:MetaAtom}.

As for the characterization results presented with respect to the MS-level designs in Sec.\ \ref{Sec:MSLevel}, we first shift the reference plane of the incident and reflected waves to $z\to 0^{-}$ and that of the transmitted wave to $z\to 0^{+}$, so as to comply with the meta-atom definition of the scattering coefficients in Eqs.\ (\ref{Eq:Ey}), (\ref{Eq:scatTE}), and (\ref{Eq:TECoeff}); hence, the relevant reflection and transmission coefficients at the interrogation angle $\theta_{\mathrm{c,mm}}^{zz}$ are modified into \cite{Pozar2012}
\begin{equation}
\label{Eq:PlaneShift}
    \begin{aligned}        
        r_{0^{-}\to0^{-}}\left(\theta_{\mathrm{c,mm}}^{zz}\right)&=r_{-\frac{d}{2}\to -\frac{d}{2}}\left(\theta_{\mathrm{c,mm}}^{zz}\right)e^{jk_{0}d\cos\theta_{\mathrm{c,mm}}^{zz}},\\
        t_{0^{-}\to0^{+}}\left(\theta_{\mathrm{c,mm}}^{zz}\right)&=t_{-\frac{d}{2}\to \frac{d}{2}}\left(\theta_{\mathrm{c,mm}}^{zz}\right)e^{jk_{0}d\cos\theta_{\mathrm{c,mm}}^{zz}},
    \end{aligned}
\end{equation}
where the scattering coefficients for the PCB MS design in Sec. \ref{Sec:MSLevel}, $r_{-\frac{d}{2}\to -\frac{d}{2}}\left(\theta_{\mathrm{c,mm}}^{zz}\right)$ and $t_{-\frac{d}{2}\to \frac{d}{2}}\left(\theta_{\mathrm{c,mm}}^{zz}\right)$, are defined in Eq.\ (\ref{Eq:RT_Cascade}) for the original reference planes of $z=\pm\frac{d}{2}$. Next, we use these effective values of scattering coefficients, $r_{0^{-}\to0^{-}}\left(\theta_{\mathrm{c,mm}}^{zz}\right)$ and $t_{0^{-}\to0^{+}}\left(\theta_{\mathrm{c,mm}}^{zz}\right)$, instead of $r^{\mathrm{TE}}\left(\theta_{\mathrm{c,mm}}^{zz}\right)$ and $t^{\mathrm{TE}}\left(\theta_{\mathrm{c,mm}}^{zz}\right)$, in Eqs.\ (\ref{Eq:CharTangential}) and (\ref{Eq:CharNormal}) with $\theta_{\mathrm{c,mm}}^{zz}=50^{\circ}$. We thus obtain the results presented in Sec.\ \ref{Sec:MSLevel}.

\begin{figure}
    \includegraphics[width=0.42\textwidth]{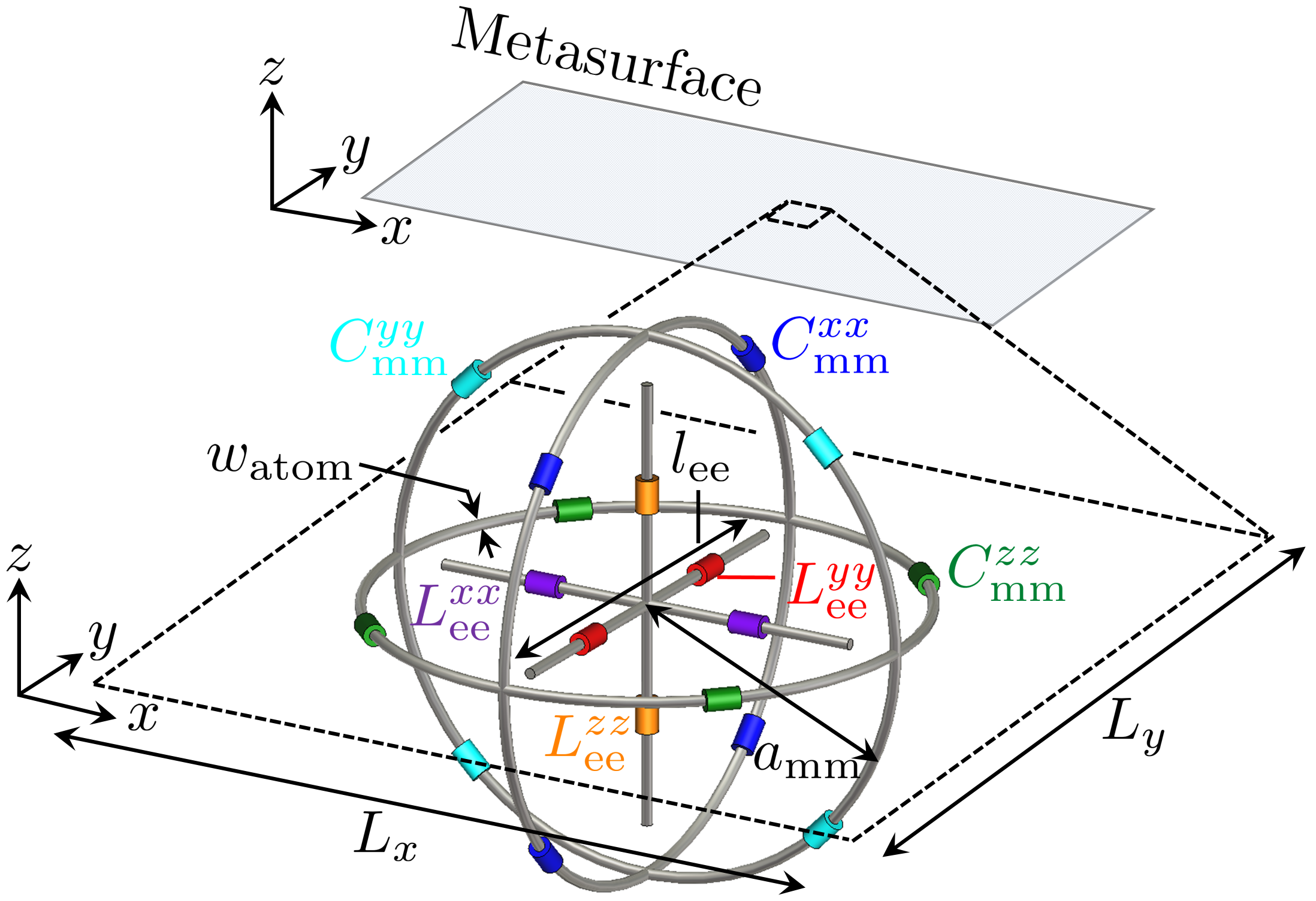}
    \caption{Physical meta-atom configuration for realizing the GHC for both TE and TM polarizations [Eqs.\ (\ref{Eq:GenHuygensTE}) and (\ref{Eq:GenHuygensTM})] at $f=20$ GHz. $L_{x}=L_{y}=3$ mm $\approx0.2\lambda_0$ are the unit-cell sizes along $x$ and $y$, respectively; $w_{\mathrm{atom}}=0.03$ mm $\approx0.002\lambda_0$ is the wire diameter common to all inclusions; $a_{\mathrm{mm}}=0.75$ mm $\approx$ $0.05 \lambda_0$ is the outer radius common to all the loops; $l_{\mathrm{ee}}=2\times 0.6$ mm $\approx 2\times 0.04\lambda_0$ is the total length of each straight inductively loaded PEC wire. Each of the $\widetilde{\chi}_{\mathrm{mm}}^{xx}$, $\widetilde{\chi}_{\mathrm{mm}}^{yy}$, and $\widetilde{\chi}_{\mathrm{mm}}^{zz}$ loops is loaded by four lumped capacitive loads of $C_{\mathrm{mm}}^{xx}$ (blue), $C_{\mathrm{mm}}^{yy}$ (cyan), and $C_{\mathrm{mm}}^{zz}$ (green) capacitance each, respectively; similarly, the straight $\widetilde{\chi}_{\mathrm{ee}}^{xx}$, $\widetilde{\chi}_{\mathrm{ee}}^{yy}$, and $\widetilde{\chi}_{\mathrm{ee}}^{zz}$ wires are loaded by two inductive loads of $L_{\mathrm{ee}}^{xx}$, $L_{\mathrm{ee}}^{yy}$, and $L_{\mathrm{ee}}^{zz}$ inductance each, respectively. Designed load values for the demonstration herein are detaied in Appendix \ref{App:AllPolarizations}.}
\label{Fig:CompMA}
\end{figure}

\section{Generalization of the GHC meta-atom design supporting both TE and TM polarizations}
\label{App:AllPolarizations}
To generalize the meta-atom concept presented in Sec.\ \ref{Sec:MetaAtom} and Fig.\ \ref{Fig:MetaAtom} and thus support the GHC for both the TE and TM polarizations at once [Eqs.\ (\ref{Eq:GenHuygensTE}) and (\ref{Eq:GenHuygensTM})], we propose the extended configuration in Fig.\ \ref{Fig:CompMA} for $f=20$ GHz: three perpendicular capacitively loaded PEC loops and inductively loaded straight PEC wires, each of which manifests its respective magnetic ($\widetilde{\chi}_{\mathrm{mm}}^{xx}$, $\widetilde{\chi}_{\mathrm{mm}}^{yy}$, and $\widetilde{\chi}_{\mathrm{mm}}^{zz}$) or electric ($\widetilde{\chi}_{\mathrm{ee}}^{xx}$, $\widetilde{\chi}_{\mathrm{ee}}^{yy}$, and $\widetilde{\chi}_{\mathrm{ee}}^{zz}$) susceptibility component in the overall response, are concentrically placed in a deep-subwavelength unit-cell of $L_{x}\times L_{y}=3$ mm $\times$ $3$ mm $\approx0.2\lambda_0\times0.2\lambda_0$ size.

Similarly to the design procedure in Sec.\ \ref{Sec:MetaAtom} and Appendix \ref{App:Char}, we set subwavelength geometrical properties for the inclusions (detailed in the caption of Fig.\ \ref{Fig:CompMA}) and tune the capacitive and inductive loads, $C_{\mathrm{mm}}^{xx,yy,zz}$ and $L_{\mathrm{ee}}^{xx,yy,zz}$ to accomplish the GHC of Eqs.\ (\ref{Eq:GenHuygensTE}) and (\ref{Eq:GenHuygensTM}), focusing on an exemplary set of $\widetilde{\chi}_{\mathrm{GHC}}^{\mathrm{TE}}=\widetilde{\chi}_{\mathrm{GHC}}^{\mathrm{TM}}=-0.5$ goal values for the following demonstration. Note that in general, the range of achievable susceptibility values for the enhanced meta-atom configuration in Fig.\ \ref{Fig:CompMA} is not limited to these specific values: in principle, one may judiciously select different values for the load parameters and achieve any arbitrary independent (real) values of $\widetilde{\chi}_{\mathrm{GHC}}^{\mathrm{TE}}$ and $\widetilde{\chi}_{\mathrm{GHC}}^{\mathrm{TE}}$. As before, due to the high symmetry of the meta-atom geometry, the loop and wire inclusions barely interact with one another, such that each susceptibility component can be individually tuned with negligible influence on the other responses.

We first establish LUTs for each of the magnetic inclusions (not shown), which associate each of the capacitance values, $C_{\mathrm{mm}}^{xx}$, $C_{\mathrm{mm}}^{yy}$, and $C_{\mathrm{mm}}^{zz}$, to its respective susceptibility value $\widetilde{\chi}_{\mathrm{mm}}^{xx}$, $\widetilde{\chi}_{\mathrm{mm}}^{yy}$, and $\widetilde{\chi}_{\mathrm{mm}}^{zz}$; this is done by following a procedure identical to the one in Appendix \ref{App:Char}, based on \cite{Zaluski2016}. Next, we utilize these LUTs and tune $C_{\mathrm{mm}}^{xx}=C_{\mathrm{mm}}^{yy}=72.876$ fF and $C_{\mathrm{mm}}^{zz}=62.605$ fF to thus achieve $\widetilde{\chi}_{\mathrm{mm}}^{xx}=-\widetilde{\chi}_{\mathrm{mm}}^{zz}=\widetilde{\chi}_{\mathrm{GHC}}^{\mathrm{TE}}=-0.5$ and $\widetilde{\chi}_{\mathrm{mm}}^{yy}=\widetilde{\chi}_{\mathrm{GHC}}^{\mathrm{TM}}=-0.5$, as requested by the GHC of Eqs.\ (\ref{Eq:GenHuygensTE}) and (\ref{Eq:GenHuygensTM}) with the designated goal values above.

\begin{figure}
    \includegraphics[width=0.48\textwidth]{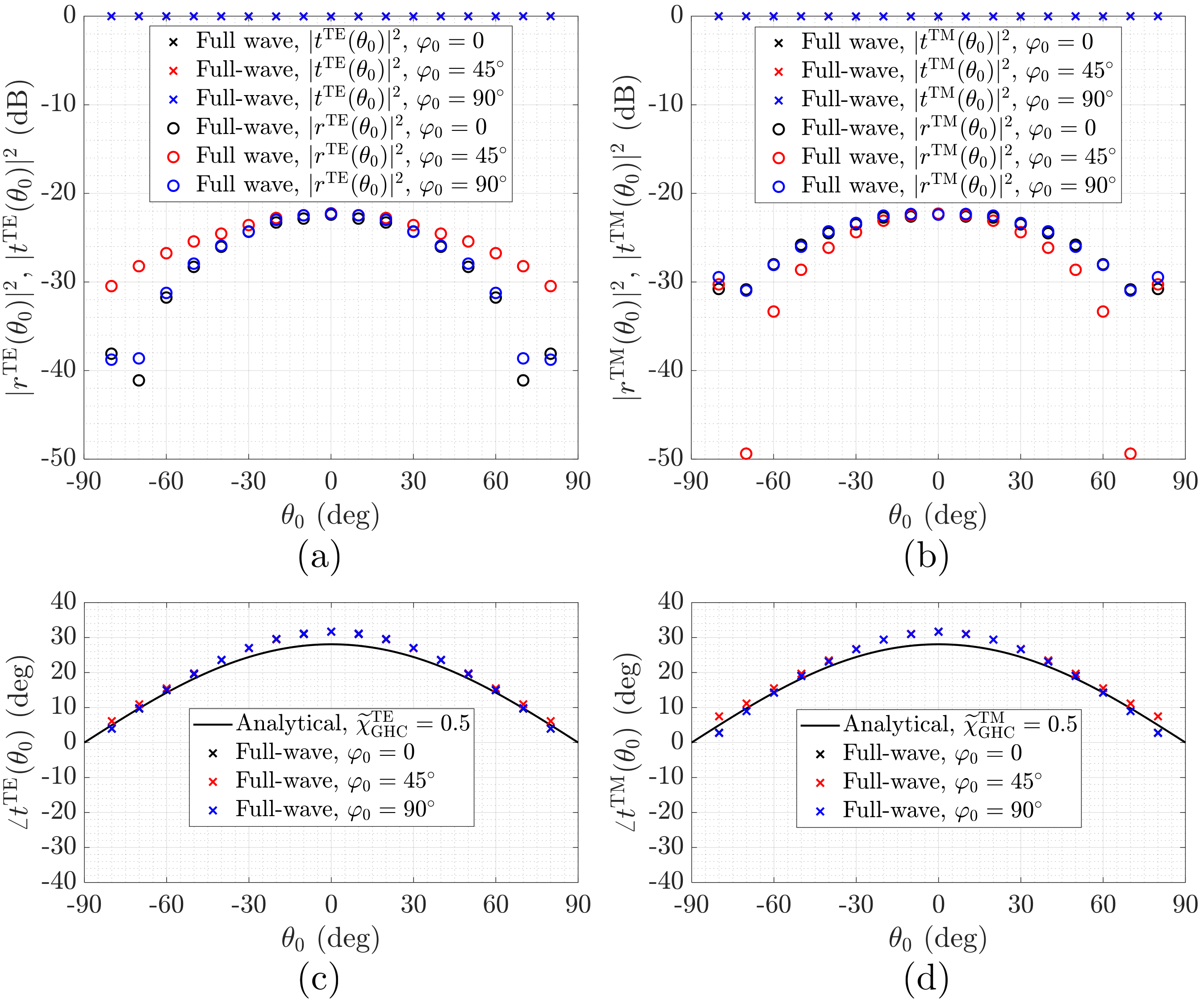}
\caption{Scattering TE [(a) and (c)] and TM [(b) and (d)] results for the $\widetilde{\chi}_{\mathrm{GHC}}^{\mathrm{TE}}=\widetilde{\chi}_{\mathrm{GHC}}^{\mathrm{TE}}=-0.5$ meta-atom in Fig.\ \ref{Fig:CompMA}: full-wave transmittance [$\times$ markers in (a) and (b)], reflectance [$\circ$ markers in (a) and (b)], and transmission phase [$\times$ markers in (c) and (d)] vs.\ elevation $\theta_0$ for azimuth values of $\varphi_0=0$ (black), $\varphi_0=45^{\circ}$ (red), and $\varphi_0=90^{\circ}$ in comparison to analytical predictions (solid lines).}
\label{Fig:CompMAresults}
\end{figure}

\begin{figure*}
    \includegraphics[width=\textwidth]{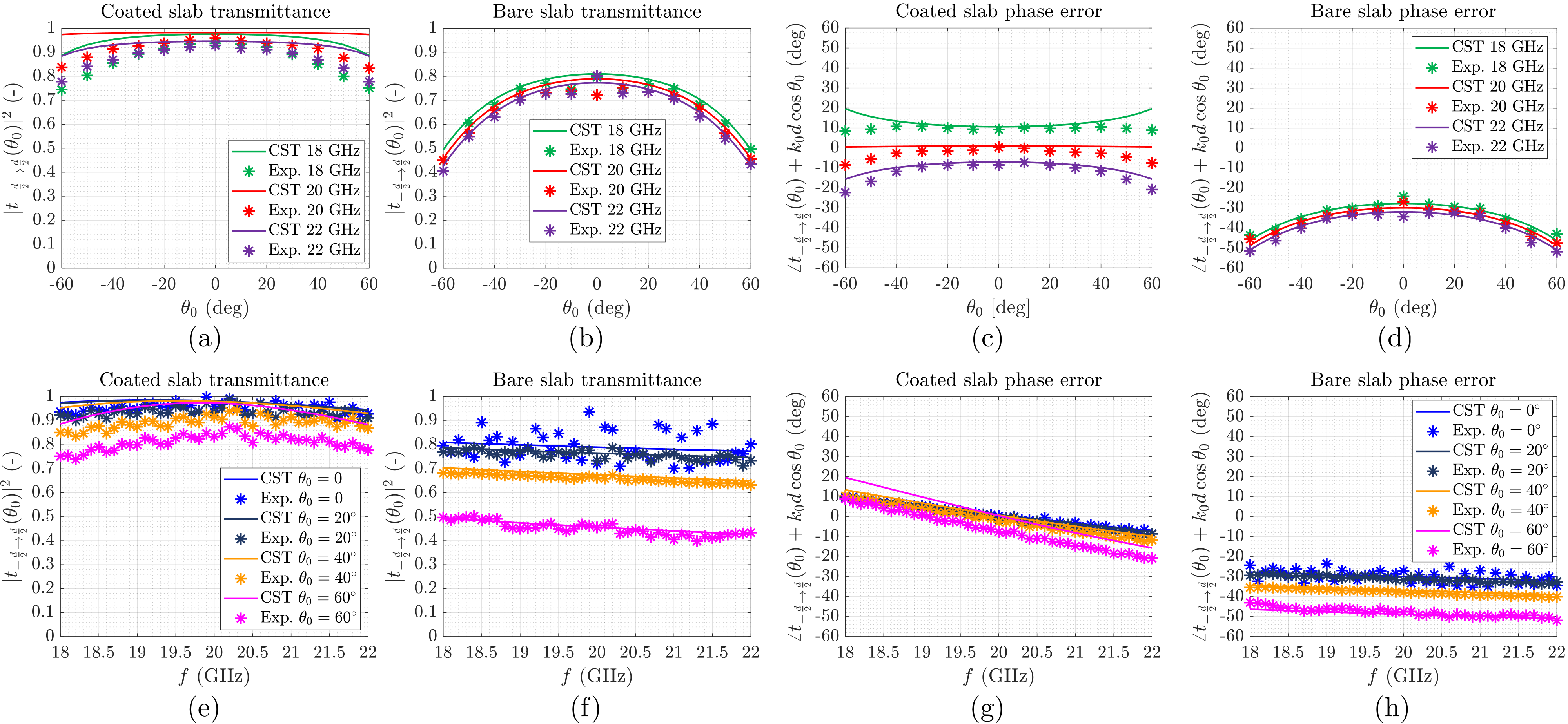}
\caption{(a)--(d) Angular dependence of the full-wave (solid lines, CST) and experimentally measured ($*$ markers) transmittance [(a) and (b)] and phase error [(c) and (d)] for the coated [(a) and (c)] and bare [(b) and (d)] slabs at $f=18$ GHz (green), $f=20$ GHz (red), and $f=22$ GHz (purple). (e)--(h) Frequency dependence of the full-wave (solid lines) and experimentally measured ($*$ markers) transmittance [(e) and (f)] and phase error [(g) and (h)] for the coated [(e) and (g)] and bare [(f) and (h)] slabs at $\theta_0=0$ (blue), $\theta_0=20^{\circ}$ (dark navy blue), $\theta_0=40^{\circ}$ (orange), $\theta_0=60^{\circ}$ (magenta).}
\label{Fig:ExpFreq}
\end{figure*}

Subsequently, to reduce the amount of required simulations, we follow symmetry considerations (ensued by the meta-atom geometry and the chosen in-plane isotropic set of goal values, i.e., $\widetilde{\chi}_{\mathrm{GHC}}^{\mathrm{TE}}=\widetilde{\chi}_{\mathrm{GHC}}^{\mathrm{TM}}$) to set $L_{\mathrm{ee}}^{xx}=L_{\mathrm{ee}}^{yy}$ and sweep their common value while monitoring the plane-wave reflectance for the TE polarization at the near-grazing angle $\theta_0=85^{\circ}$ in CST (under periodic boundary conditions). We find that the values $L_{\mathrm{ee}}^{xx}=L_{\mathrm{ee}}^{yy}=8.3672$ nH achieve vanishing reflectance at this interrogation angle, and, hence, fix them to thus guarantee $\widetilde{\chi}_{\mathrm{ee}}^{yy}=\widetilde{\chi}_{\mathrm{GHC}}^{\mathrm{TE}}=-0.5$ and
$\widetilde{\chi}_{\mathrm{ee}}^{xx}=\widetilde{\chi}_{\mathrm{GHC}}^{\mathrm{TM}}=-0.5$, as desired. Next, we sweep the value of $L_{\mathrm{ee}}^{zz}$ while monitoring the TM plane-wave reflectance at $\theta_0=85^{\circ}$; we fix the value $L_{\mathrm{ee}}^{zz}=8.323$ nH, which leads to zero reflectance at this angle and guarantees $\widetilde{\chi}_{\mathrm{ee}}^{zz}=-\widetilde{\chi}_{\mathrm{GHC}}^{\mathrm{TM}}=0.5$, and conclude the design procedure.

To show that the finalized meta-atom in Fig.\ \ref{Fig:CompMA} indeed fully meets the GHC as intended, we probe its plane-wave scattering properties for both polarizations at $-80^{\circ}<\theta_0<80^{\circ}$ (under periodic boundary conditions). The full-wave TE [black markers in Fig.\ \ref{Fig:CompMAresults}(a)] and TM [black markers in Fig.\ \ref{Fig:CompMAresults}(b)] reflectance results ($|r^{\mathrm{TE}}\left(\theta_0\right)|^2$ and $|r^{\mathrm{TM}}\left(\theta_0\right)|^2$ in dB units, $\circ$ markers) exhibit excellent performance, remaining below $0.6\%\approx-22.3$ dB; accordingly, the TE and TM transmittance ($|t^{\mathrm{TE}}\left(\theta_0\right)|^2$ and $|t^{\mathrm{TM}}\left(\theta_0\right)|^2$ in dB, black $\times$ markers) results complement their respective reflectance values to unity, achieving similar remarkable performance and concordance with theory. Furthermore, the full-wave transmission phase [black $\times$ markers in Figs.\ \ref{Fig:CompMAresults}(c) and (d)] excellently agrees with theoretical predictions (solid lines) for both TE [Eq.\ (\ref{Eq:scatTE}), Fig.\ \ref{Fig:CompMAresults}(c)] and TM [Eq.\ (\ref{Eq:scatTM}), Fig.\ \ref{Fig:CompMAresults}(d)] polarizations as well. This indeed verifies the generalized Huygens' functionality of the meta-atom for both polarizations.

Seeing as the meta-atom configuration above is transversely isotropic ($\widetilde{\chi}_{\mathrm{ee}}^{xx}=\widetilde{\chi}_{\mathrm{ee}}^{yy}$ and $\widetilde{\chi}_{\mathrm{mm}}^{xx}=\widetilde{\chi}_{\mathrm{mm}}^{yy}$, applicable to the particular scenario of $\widetilde{\chi}_{\mathrm{GHC}}^{\mathrm{TE}}=\widetilde{\chi}_{\mathrm{GHC}}^{\mathrm{TM}}$), we expect its scattering properties at a fixed elevation angle of incidence $\theta_0$ to be invariant to the azimuth $\varphi_0$ of the plane of incidence. This hypothesis is indeed confirmed by comparing the scattering results for the exemplary azimuth angles (defined, as standard, relative to the $x$ axis) of $\varphi_0=45^{\circ}$ (red) and $\varphi_0=90^{\circ}$ (blue)  to those of the original $xz$-plane of incidence, $\varphi_0=0$ (black) in Fig.\ \ref{Fig:CompMAresults}; indeed they all practically coincide, as anticipated. These results strongly validate the GHC for both polarizations and, specifically to in-plane isotropic designs, for all planes of incidence, showing that this powerful concept can be systematically achieved by means of physical configurations.

\section{Reflectance and absorption calculation for the practical line-source illumination}
\label{App:RefAbsorbCalc}
Herein, we describe in more detail the numerical reflectance and absorption calculation performed for the results in Fig.\ \ref{Fig:PCBFields}. The reflectance calculation is done in several steps: first, we numerically integrate the $z$-directed time-averaged Poynting vector, as provided by CST, slightly below the MS at $z=-\frac{d}{2}-t_{\mathrm{copper}}=-0.78$ mm ($t_{\mathrm{copper}}=0.018$ is the standard commercial copper thickness selected for the specimens herein, i.e., 0.5 oz.\ per square foot), denoted by $P_{\mathrm{up}}$. This quantity approximately represents the difference between the incident power and the reflected power (for the fixed excitation current $I_0$). Next, subject to the same excitation, we numerically integrate the $z$-directed time-averaged power flowing below the dipole on the $z=-\frac{2\lambda_0}{3}+\frac{d}{2}+t_{\mathrm{copper}}=-9.22$ mm plane (which is far from the dipole by the same distance of the previous plane), denoted by $P_{\mathrm{down}}$. This value approximately accounts for the sum of the incident and reflected power (in negative sign, as the total power flows downwards to the $-z$ direction). Thus the reflectance, i.e., the ratio between reflected and incident power, is estimated via $\frac{1+P_{\mathrm{up}}/P_{\mathrm{down}}}{1-P_{\mathrm{up}}/P_{\mathrm{down}}}$ (similar to standing-wave ratio, SWR, calculations \cite{Pozar2012}). This calculation is individually applied for the coated and bare slabs with the line-source excitation in Fig.\ \ref{Fig:PCBFields}.

The absorption (mostly due to the copper traces, as reported by CST) is calculated for the entire structure in CST (internal feature). Next, this absorbed power is divided by the incident power, which is estimated via the previous calculations through $\frac{P_{\mathrm{up}}-P_{\mathrm{down}}}{2}$. Again, this calculation is individually applied for the coated and bare slabs in Fig.\ \ref{Fig:PCBFields}. This finalizes the calculation and establishes an estimate for the absorption as well.

\section{Detailed investigation of frequency response for the MS-level design}
\label{App:Freq}

To further explore the frequency response of the devised prototype in Fig.\ \ref{Fig:PCBConfig}(c) and Fig.\ \ref{Fig:ExpSetup}, we present the full-wave (solid lines) and experimental ($*$ markers) results for the coated slab's transmittance [Fig.\ \ref{Fig:ExpFreq}(a)] and phase error [Fig.\ \ref{Fig:ExpFreq}(c)] vs.\ angle at several representative frequencies; similar reference plots are shown in Figs.\ \ref{Fig:ExpFreq}(b) and (d) for the bare slab. Indeed, the measured phase error for the coated substrate is in very good agreement with the full-wave results, especially at the range $-40^{\circ}\leq\theta_0\leq40^{\circ}$, and the angular behavior of the measured transmittance follows the trend of the full-wave results: the transmittance slightly drops when the frequency deviates from the intended frequency of operation ($f=20$ GHz), especially for oblique angles.

The measured ($*$ markers) and full-wave (solid lines) frequency response of the transmittance [Figs.\ \ref{Fig:ExpFreq}(e) and (f)] and phase error [Figs.\ \ref{Fig:ExpFreq}(g) and (h)] are plotted for several selected angles for both the coated [Figs.\ \ref{Fig:ExpFreq}(e) and (g)] and bare [Figs.\ \ref{Fig:ExpFreq}(f) and (h)] slabs. For the coated slab, the measured trends of frequency dependence are highly congruent with those of the full-wave ones: the phase error monotonically decreases as a function of frequency with matching slopes for all angles, whereas the measured transmittance peaks for all angles, at a slightly shifted frequency of $20.2$ GHz ($1\%$ frequency error relative to the goal frequency of $20$ GHz), while the transmittance bandwidth decreases as the incident angle is made more oblique. Considering the minor frequency shift to $20.2$ GHz the measured power transmittance over all the range of $-60^{\circ}\leq\theta_0\leq60^{\circ}$ is even further improved to more than $87.5\%$ (compared to $83.7\%$ at 20 GHz).


%

\end{document}